\documentclass[twocolumn,english,aps,superscriptaddress,prl,showpacs]{revtex4}
\usepackage[T1]{fontenc}
\usepackage[latin9]{inputenc}
\usepackage{amsmath}
\usepackage{amssymb}
\usepackage{graphicx}
\usepackage{esint}
\usepackage{color}

\makeatletter
\@ifundefined{textcolor}{}
{%
 \definecolor{BLACK}{gray}{0}
 \definecolor{WHITE}{gray}{1}
 \definecolor{RED}{rgb}{1,0,0}
 \definecolor{GREEN}{rgb}{0,1,0}
 \definecolor{BLUE}{rgb}{0,0,1}
 \definecolor{CYAN}{cmyk}{1,0,0,0}
 \definecolor{MAGENTA}{cmyk}{0,1,0,0}
 \definecolor{YELLOW}{cmyk}{0,0,1,0}
}

\newcommand{\aualyb}{\mbox{Au}_{51}\mbox{Al}_{34}\mbox{Yb}_{15}}
\newcommand{\tk}{T_{\rm K}}
\newcommand{\tkt}{T_{\rm K}^{\rm typ}}
\newcommand{\tkm}{T_{\rm K}^{\rm min}}

\makeatother

\usepackage{babel}
\begin{document}

\title{Non-Fermi-Liquid Behavior in Metallic Quasicrystals with Local Magnetic
Moments}

\author{Eric C. Andrade}

\affiliation{Instituto de F\'{i}sica Te\'{o}rica, Universidade Estadual Paulista, Rua
Dr. Bento Teobaldo Ferraz, 271 - Bloco II, 01140-070, S\~{a}o Paulo, SP,
Brazil}

\author{Anuradha Jagannathan}

\affiliation{Laboratoire de Physique des Solides, CNRS-UMR 8502, Universit\'{e} Paris-Sud,
91405 Orsay, France}

\author{Eduardo Miranda}

\affiliation{Instituto de F\'{\i}sica Gleb Wataghin, Unicamp, Rua S\'{e}rgio Buarque de Holanda, 777, CEP 13083-859 Campinas, SP, Brazil}

\author{Matthias Vojta}

\affiliation{Institut f\"{u}r Theoretische Physik, Technische Universit\"{a}t Dresden,
01062 Dresden, Germany}

\author{Vladimir Dobrosavljevi\'{c}}

\affiliation{Department of Physics and National High Magnetic Field Laboratory,
Florida State University, Tallahassee, Florida 32306, USA}

%%%%%%%%%%%%%%%%%%%%%%%%%%%%%%%%%%%%%%%%%%%%%%%%%%%%%%%%%%%%%%%%%%%%%%%

\begin{abstract}

Motivated by the intrinsic non-Fermi-liquid behavior observed in the
heavy-fermion quasicrystal $\aualyb$, we study the low-temperature
behavior of dilute magnetic impurities placed in metallic quasicrystals.
We find that a large fraction of the magnetic moments are not quenched
down to very low temperatures $T$, leading to a power-law distribution
of Kondo temperatures $P(\tk)\sim\tk^{\alpha-1}$, with a non-universal
exponent $\alpha$, in a remarkable similarity to the Kondo-disorder
scenario found in disordered heavy-fermion metals. For $\alpha<1$,
the resulting singular $P(\tk)$ induces non-Fermi-liquid behavior
with diverging thermodynamic responses as $T\rightarrow0$.

\end{abstract}

\date{\today}


\maketitle

%%%%%%%%%%%%%%%%%%%%%%%%%%%%%%%%%%%%%%%%%%%%%%%%%%%%%%%%%%%%%%%%%%%%%%%

\emph{Introduction.}---Fermi-liquid (FL) theory forms the basis of our
understanding of interacting fermions. It works in a broad range of
systems, from weakly correlated metals \cite{ashcroft} to strongly
interacting heavy fermions \cite{Hewson_kondo}. Over the past decades,
however, the properties of numerous metals have been experimentally
found to deviate from FL predictions \cite{stewart01,maple10}, and
much effort has been devoted to the understanding of such non-Fermi-liquid
(NFL) behavior. One interesting avenue is provided by quantum critical
points (QCPs): NFL physics may occur in the associated quantum critical
regime which is reached upon tuning the system via a nonthermal control
parameter such as pressure, doping, or magnetic field \cite{coleman01,lrvw07}.

Remarkably, recent experiments have provided compelling evidence of
NFL behavior without fine-tuning in the heavy-fermion \emph{quasicrystal}
$\aualyb$ \cite{deguchi12,watanuki12}. Furthermore, Ref.~\onlinecite{deguchi12}
also reports that no NFL behavior emerges when one considers a crystalline
approximant instead of the quasicrystal, suggesting that this NFL
regime is associated with the particular electronic states present
in the quasicrystal but not in the approximant \cite{kohmoto87,tsunetsugu91a,yuan00,grimm03,anu06,deLa14}.
Conventional QCP approaches have been employed to explain the fascinating
behavior in this alloy \cite{watanabe13,shaginyan13}, but they consider
the effects of quasicrystalline environment of the conduction electrons
only minimally.

In this work we intend to close this gap by presenting a detailed
calculation of the fate of isolated localized magnetic moments when
placed in both two- and three-dimensional quasicrystals. Our
results for dilute impurities show that a considerable fraction of
impurity moments is not quenched down to very low temperatures, leading
to a power-law distribution of Kondo temperatures, $P(\tk)\propto\tk^{\alpha-1}$,
with a nonuniversal exponent $\alpha$. This results in NFL behavior
in both $\chi$ and $C/T$ as $T\rightarrow0$: $\chi\sim C/T\sim T^{\alpha-1}$
\cite{nfl_review05}, a scenario very reminiscent of the Kondo effect
in disordered metals \cite{vlad92a,eduardo96,cornaglia06,kettemann09,kettemann12,miranda14}.
Moreover, we show that the strong energy dependence of the electronic
density of states (DOS) characteristic of a quasicrystal leads to
a situation such that small changes in the model parameters (band
filling, Kondo coupling, etc.) may drive the system in and out of the
NFL region. %%%% a fact that may help understand conflicting results for the  quasicrystal and its crystalline approximant. %%%%

%%%%%%%%%%%%%%%%%%%%%%%%%%%%%%%%%%%%%%%%%%%%%%%%%%%%%%%%%%%%%%%%%%%%%%%

\emph{Quasicrystalline wave functions.}---A quasicrystal exhibits a small set of local environments, which reappear again and again,
albeit not in a periodic fashion. Their pattern is not random either,
since the structure factor shows sharp Bragg peaks, although their
symmetry is noncrystallographic \cite{shechtman84}. The $n$-fold
symmetries (with values of $n=5,8,10,\ldots$) seen in the diffraction
pattern of quasicrystals arise due to the fact that the local environments
occur with $n$ equiprobable orientations.

The structure factor of quasicrystals is densely filled in reciprocal
space with diffraction spots \cite{shechtman84} of widely differing
intensities. The brighter peaks are expected to lead to strong scattering
of conduction electrons, giving rise to spikes in the DOS \cite{smith87,zijlstra00}.
The scattering due to the remaining peaks, while weaker, results in
wave functions which show fluctuations at all length scales. %%% The resulting multifractal character of wavefunctions is a consequence of the invariance of quasicrystals under scale transformations, called inflation-deflation symmetry. %%%
The Fibonacci chain, a one-dimensional quasicrystal, provides an example
of such wave functions \cite{kohmoto87}, often referred to as \emph{critical}
\cite{kohmoto87,tsunetsugu91a,yuan00,grimm03,anu06}, in analogy with
those found at the Anderson metal-insulator transition \cite{richardella10,rodriguez10}.%%% Numerical studies of tight-binding models in higher dimensional quasicrystals indicate that wavefunctions also have multifractal scaling properties in D\ge2. In other words, wavefunctions in a quasicrystal, generically, are intermediate character between localized and extended states, and are characterized by power law singularities in space and in energy%%%

%%%%% Very little is known about the effects of electron-electron interactions, and the screening of impurities in a quasicrystal. This is the problem we now address, using, the example of a %%%%

%%%%%%%%%%%%%%%%%%%%%%%%%%%%%%%%%%%%%%%%%%%%%%%%%%%%%%%%%%%%%%%%%%%%%%%

\emph{Tiling model.}--- For simplicity, we consider models on quasiperiodic
tilings. We first report results obtained for a $2D$ tiling, where
it is easier to handle large system sizes numerically. 
In the Supplemental Material~\cite{suppl},
we show calculations for a $3D$ tiling \cite{lcd} with very similar results,
confirming that our scenario is independent of both tiling details and dimensionality.

\begin{figure}[t]
\begin{centering}
\includegraphics[scale=0.3]{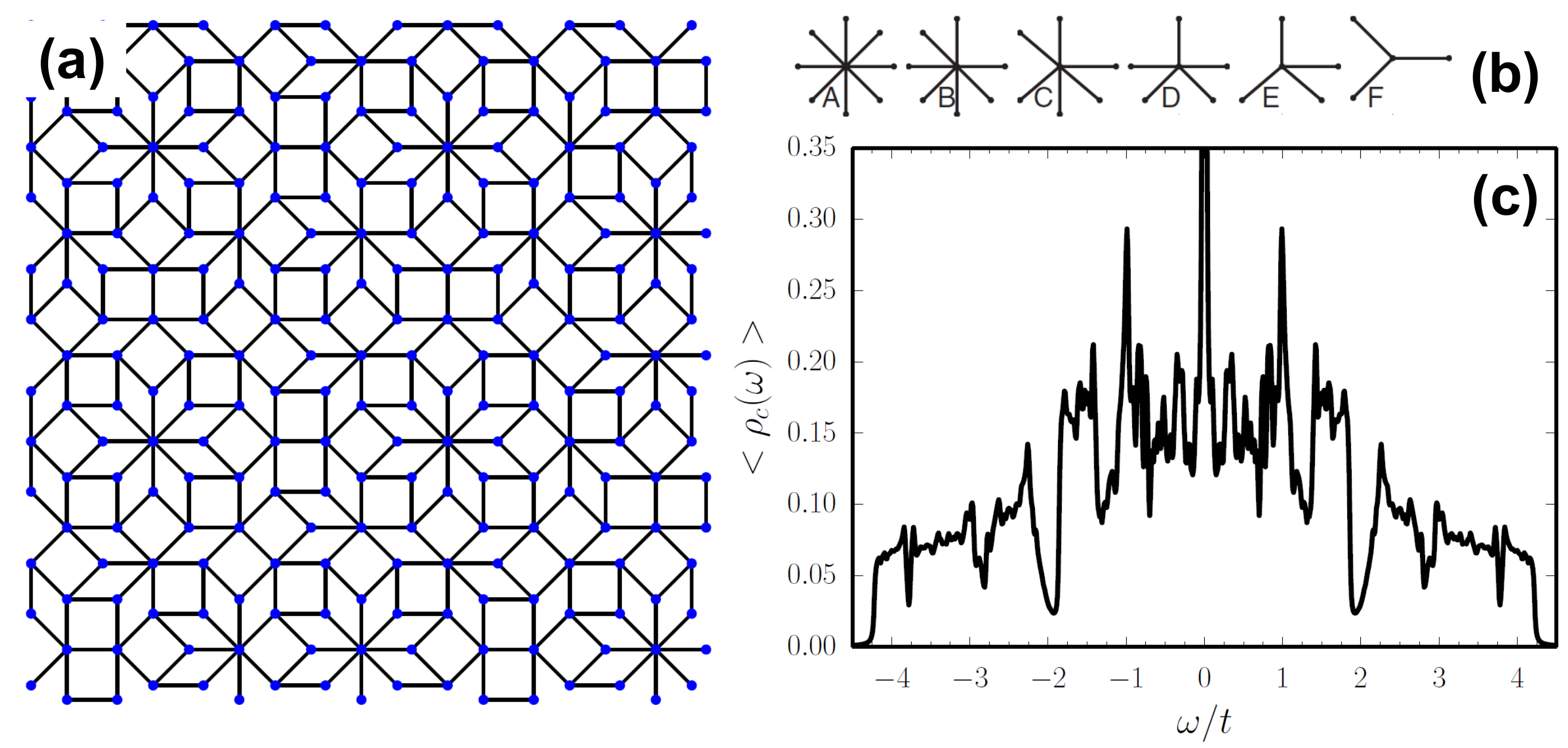} 
\par\end{centering}

\caption{\label{fig:qc} Quasicrystal geometrical and electronic properties.
(a) Square approximant for the perfect octagonal tiling with $N_{a}=239$
sites. (b) The six local site environments with $z=3,\ldots,8$ nearest
neighbors. (c) The total DOS as a function of the energy for the $N_{a}=8119$
approximant averaged over $N_{\phi}=64$ twist angles. }
\end{figure}

The $2D$ tiling we consider is the octagonal tiling (Ammann-Beenker) \cite{socolar89},
Fig.~\ref{fig:qc}(a). This tiling is composed of two types of decorated
tiles: squares and $45^{o}$ rhombuses, which combine to create six
distinct local environments with coordination number $z=3,\cdots,8$,
Fig.~\ref{fig:qc}(b). 

As a minimal model to describe the electronic properties of quasicrystals,
we consider a nearest-neighbor tight-binding Hamiltonian in standard
notation

\begin{eqnarray}
\mathcal{H}_{c} & = & -t\sum_{\left\langle ij\right\rangle ,\sigma}\left(c_{i\sigma}^{\dagger}c_{j\sigma}+c_{j\sigma}^{\dagger}c_{i\sigma}\right).\label{eq:hc}
\end{eqnarray}
%where $c_{i\sigma}^{\dagger}\left(c_{i\sigma}\right)$ is the creation
%(annihilation) operator of a $c$-electron at site $i$ with spin
%$\sigma$ and $t$ is the hopping amplitude between sites $i$ and $j$. 
In the following, energies are measured in units of $t$. In our calculation,
we consider periodic approximants of the octagonal tiling of sizes
$N_{a}=7$, $41$, $239$, $1393$, and $8119$, obtained by the standard
method of projecting down from a higher dimensional cubic lattice,
as in previous works \cite{socolar89,levine87,deneau89,benza91}.
To reduce finite-size effects we use twisted boundary conditions,
i.e., $\psi\left(\vec{r}+L\hat{x}+L\hat{y}\right)=e^{i\phi_{x}}e^{i\phi_{y}}\psi\left(\vec{r}\right)$
for a sample of linear size $L$. %%%% For free particles it implies that the allowed wavevectors are now shifted as  Periodic boundary conditions (PBC) correspond to  and the antiperiodic boundary conditions (ABC) to %%%%
Our final answer is obtained averaging over $N_{\phi}$ twist angles
\cite{gross96}. %%%%

In Fig.~\ref{fig:qc}(c) we show the well-known total DOS for the
octagonal tiling $\left\langle \rho_{c}\left(\omega\right)\right\rangle =\sum_{i=1}^{N_{a}}\rho_{i}^{c}\left(\omega\right)/N_{a}$,
with the local DOS at site $i$ given by $\rho_{i}^{c}\left(\omega\right)=\overline{\sum_{\nu}\left|\psi_{\nu}^{c}\left(i\right)\right|^{2}\delta\left(\omega-E_{\nu}^{c}\right)}$,
where $\psi_{\nu}^{c}$ is an eigenstate of $\mathcal{H}_{c}$ in
\eqref{eq:hc} with energy $E_{\nu}^{c}$ and the overline denotes
the average over boundary conditions. $\left\langle \rho_{c}\left(\omega\right)\right\rangle $
has a strong energy dependence with several spikes and a pronounced
dip at $\omega\approx\pm2.0t$. The large peak at $\omega=0$ is due
to families of strictly localized states, a consequence of the local
topology of the octagonal tiling \cite{grimm03,rieth95}. The spatial
structure of $\rho_{i}^{c}\left(\omega\right)$ is discussed in Ref.
\cite{suppl}, where we show that it is well described by a log-normal
distribution.

%%%We notice that periodic boundary conditions introduce frustration in the approximants we consider and that the particle-hole symmetric form obtained here comes form the average over TBC.%%%%

%%%%%%%%%%%%%%%%%%%%%%%%%%%%%%%%%%%%%%%%%%%%%%%%%%%%%%%%%%%%%%%%%%%%%%%
\emph{Local moments and large-$N$ solution.}---%%%% Given all these novel properties of electronic wave functions in quasicrystals and motivated then by the NFL observed in heavy-fermion quasicrystal %%%%
We now move to the main topic of this Letter: the investigation of
the single-impurity Kondo effect in a metallic quasicrystal. %%%%% Because in the experiments the  ions are found to be in a mixed-valence state %%%%%
Specifically, we consider the $U\rightarrow\infty$ Anderson impurity
model 
\begin{eqnarray}
\mathcal{H} & = & \mathcal{H}_{c}+E_{f}\sum_{\sigma}n_{f\sigma}+V\sum_{\sigma}\left(f_{\ell\sigma}^{\dagger}c_{\ell\sigma}+c_{\ell\sigma}^{\dagger}f_{\ell\sigma}\right).\label{eq:aim}
\end{eqnarray}
This model describes a band of noninteracting electrons ($c$ band)
which hybridize with a localized $f$ orbital located at site $\ell$.
The operator $f_{\ell\sigma}^{\dagger}\left(f_{\ell\sigma}\right)$
creates (destroys) an electron with spin $\sigma$ at the impurity
site $\ell$ and the $U\rightarrow\infty$ limit imposes the constraint
$n_{f\sigma}=f_{\ell\sigma}^{\dagger}f_{\ell\sigma}\le1$. $E_{f}$
is the $f$-level energy, measured with respect to the chemical potential
$\mu$, and the hybridization $V$ couples the impurity site to the
conduction band. To obtain quantitative results, we now turn to a
large-$N$ limit of Eq. \eqref{eq:aim} that allows us to access arbitrary
values of the model parameters \cite{readnewns83a,readnewns83b,coleman84}.
It introduces two variational parameters $Z_{\ell}$ (quasiparticle
weights) and $\tilde{\varepsilon}_{f\ell}$ (renormalized $f$-energy
levels), which are site dependent in the case of a quasicrystal. These
parameters are determined by minimization of the saddle-point free
energy (see \cite{suppl} for further details)

\begin{eqnarray}
F_{MF}^{\ell} & = & \frac{2}{\pi}\int_{-\infty}^{+\infty}f\left(\omega\right)\mbox{Im}\left[\mbox{ln}\left[\tilde{G}_{\ell}^{f}\left(\omega\right)\right]\right]d\omega\nonumber \\
 & + & \left(\tilde{\varepsilon}_{f\ell}-E_{f}\right)\left(Z_{\ell}-1\right),\label{eq:F_MF}
\end{eqnarray}
where $f\left(\omega\right)$ is the Fermi-Dirac distribution function.
The quasiparticle $f$-level Green's function is given by $\tilde{G}_{\ell}^{f}\left(\omega\right)=\left[\omega-\tilde{\varepsilon}_{f}-Z_{\ell}\Delta_{f\ell}\left(\omega\right)\right]^{-1}$,
with the $f$-electron hybridization function given by $\Delta_{f\ell}\left(\omega\right)=V^{2}G_{\ell\ell}^{c}\left(\omega\right)$,
where $G_{\ell\ell}^{c}\left(\omega\right)=\overline{\sum_{\nu}\left|\psi_{\nu}^{c}\left(\ell\right)\right|^{2}/\left(\omega-E_{\nu}^{c}\right)}$
is the $c$-electron Green's function. We define $\tk$ as the (half-)width
of the resonance at the Fermi level $\tk^{\ell}\equiv Z_{\ell}\mbox{Im}\left[\Delta_{f\ell}\left(0\right)\right]$
\cite{tk} and introduce the Kondo coupling $J\equiv2V^{2}/\left|E_{f}\right|$.
The $f$-level occupation is simply given by $n_{f\ell}=1-Z_{\ell}$.

Because each site in the quasicrystal ``sees'' a different environment,
encoded in $\Delta_{f\ell}\left(\omega\right)$, we numerically solve
Eq. \eqref{eq:F_MF}, at $T=0$, individually placing Kondo impurities
at all $N_{a}$ sites of the approximant. Therefore, for every single
impurity problem we obtain a different value of $\tk$, which we use
to construct the distribution of the Kondo temperatures $P(\tk)$.

%%%%%%%%%%%%%%%%%%%%%%%%%%%%%%%%%%%%%%%%%%%%%%%%%%%%%%%%%%%%%%%%%%%%%%%

\emph{Power-law distribution of Kondo temperatures.}---For Kondo impurities
placed in a disordered metal \cite{vlad92a,eduardo96,cornaglia06,kettemann09,kettemann12,miranda14}
it is well established that the distribution of Kondo temperatures
possesses a power-law tail at low $\tk$: $P(\tk)\propto\tk^{\alpha-1}$,
with a nonuniversal exponent $\alpha$ \cite{darko04}. For $\alpha<1$,
$P(\tk)$ becomes singular, and NFL behavior emerges in the system
\cite{nfl_review05,suppl}.

\begin{figure}[t]
\begin{centering}
\includegraphics[scale=0.25]{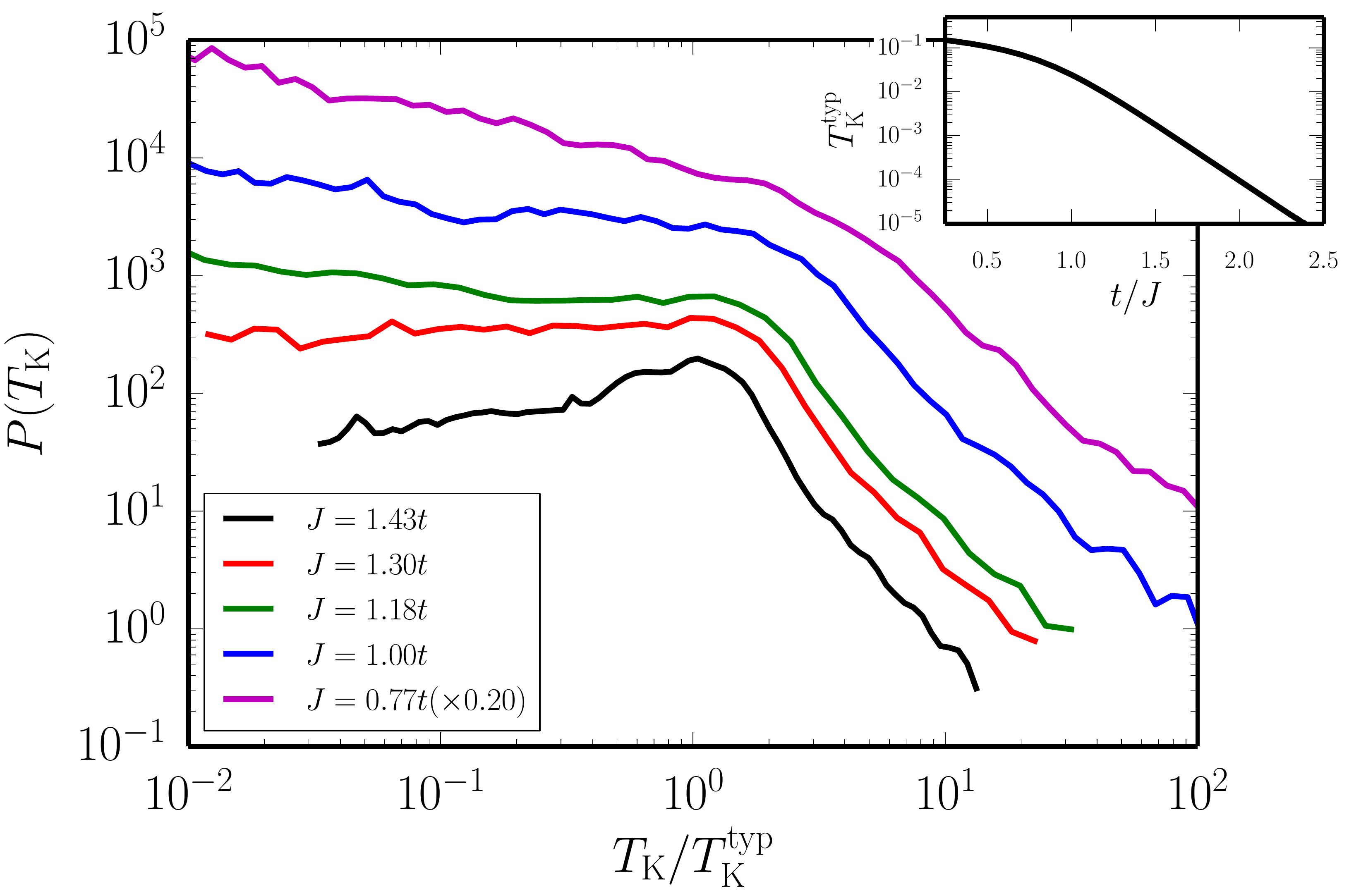} 
\par\end{centering}

\caption{\label{fig:ptk} Distribution of the local Kondo temperatures $P\left(\tk\right)$
on a log-log scale for several values of the Kondo coupling $J$;
note that the curve corresponding to $J=0.77t$ was scaled down. $\tk$
on the horizontal axis has been normalized by $\tkt$; the unrenormalized
distributions are shown in Ref.~\onlinecite{suppl}. For $\tk\lesssim\tkt$
the distributions acquire a power-law form $P\left(\tk\right)\sim\tk^{\alpha-1}$,
with the exponent $\alpha$ continuously varying with $J$. For $\alpha<1$
the distribution is singular. (Notice that for $\tk\gtrsim\tkt$,
$P\left(\tk\right)$ is also power-law like, with an exponent that
does not depend on $J$. This is \emph{not} the power-law regime we
refer to in this work). Inset: $\tkt$ as a function of $1/J$ on
a semilog scale. Here we considered $N_{a}=1393$, $\mu=-2.2t$,
and $N_{\phi}=576$. }
\end{figure}

Surprisingly, we observe the same phenomenology for quasicrystals,
with sample results shown in Fig.~\ref{fig:ptk}. Here we show the
corresponding $P(\tk)$ for the octagonal tiling at $\mu=-2.2t$ as
a function of $\tk/\tkt$ (we defined the typical value of $\tk$
as $\tkt\equiv\mbox{exp}\left[\left\langle \mbox{ln}\left(\tk\right)\right\rangle \right]$).
For approximants with $N_{a}\ge239$ a clear power-law tail emerges
for $\tk<\tkt$ with an exponent which depends on the Kondo coupling
$J$ \cite{suppl}. The dependence of $\tkt$ on $J$ is shown in
the inset of Fig.~\ref{fig:ptk}, where we see that we obtain the
expected exponential relation \cite{Hewson_kondo}.

\begin{figure}[t]
\begin{centering}
\includegraphics[scale=0.27]{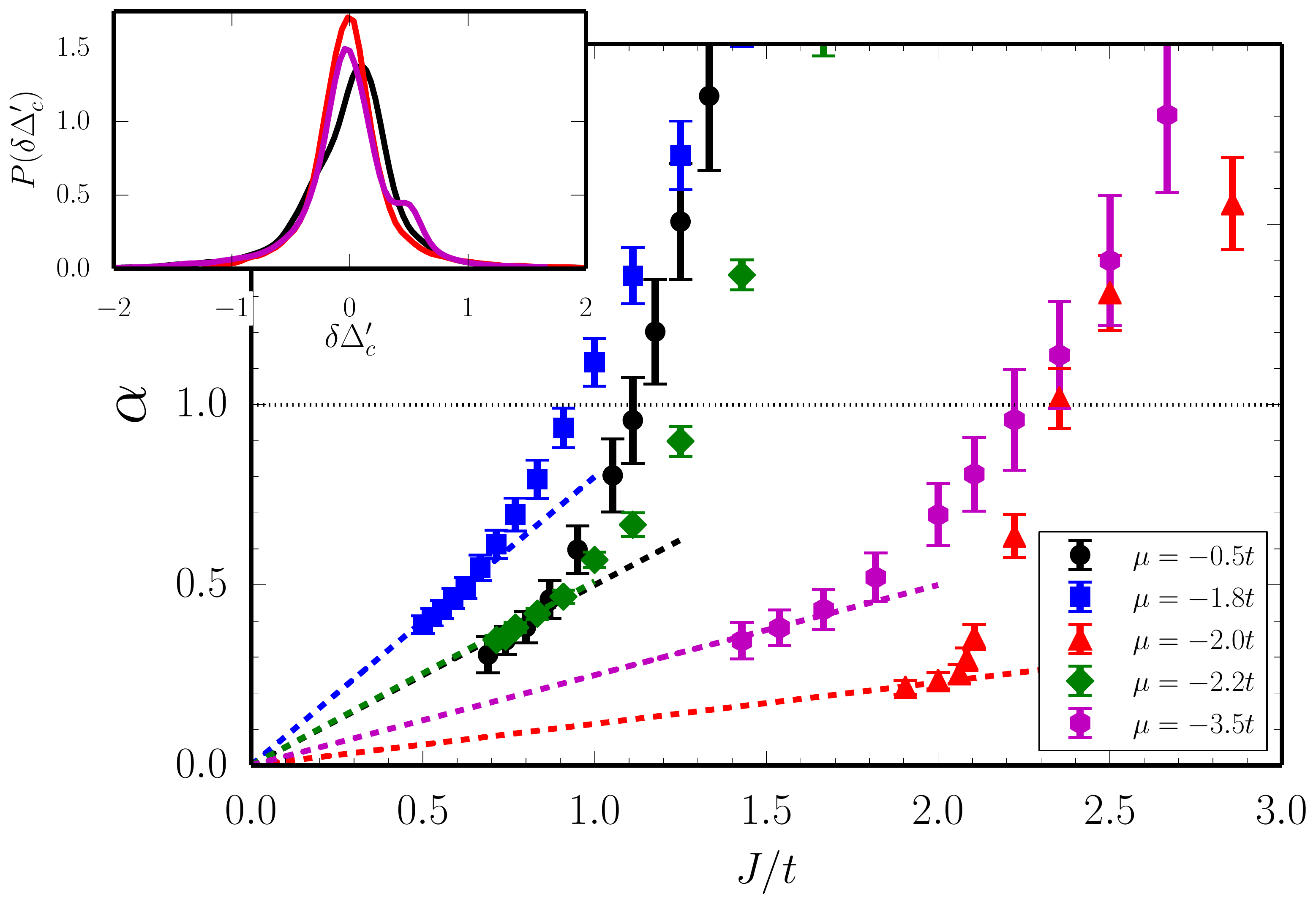} 
\par\end{centering}

\caption{\label{fig:alpha} Power-law exponent $\alpha$ as a function of the
Kondo coupling $J$ for five different positions of Fermi level $\mu$.
The dashed lines are linear fits deep into the Kondo regime where
we expect $\alpha\propto J$ to hold (see text). The horizontal dashed
line corresponds to $\alpha=1$ and marks the entrance into the NFL
region. At this point we have an average $f$-level occupation $\left\langle n_{f}\right\rangle =1-\left\langle Z\right\rangle =0.970$,
$0.995$, $0.890$, $0.995$, and $0.960$ for $\mu=-0.5t$, $-1.8t$,
$-2.0t$, $-2.2t$, and $\mu=-3.5t$, respectively. Here we considered
$N_{a}=1393$ and $N_{\phi}=576$. Inset: distribution of the real
part of the local $c$-electron cavity function fluctuations at the
Fermi level $\delta\Delta_{c}^{\prime}=\Delta_{c}^{\prime}\left(0\right)-\left\langle \Delta_{c}^{\prime}\left(0\right)\right\rangle $
for three different values of $\mu$ (the color scheme is the same
as in the main panel). Here we considered $N_{a}=8119$ and $N_{\phi}=64$. }
\end{figure}

Given the strong energy variations of $\left\langle \rho_{c}\left(\omega\right)\right\rangle $,
Fig.~\ref{fig:qc}(c), it is then natural to ask whether the form
$P(\tk)\propto\tk^{\alpha-1}$ is observed at different locations
of the Fermi level $\mu$. We checked that this is indeed the case:
in Fig.~\ref{fig:alpha} we show how the exponent $\alpha$ varies
with $J$ for several values of $\mu$ (to extract the value of $\alpha$
we followed Ref. \cite{clauset09}). The dashed straight lines correspond
to the expected behavior at low $J$ (Kondo limit) where we have $\alpha\propto J$
\cite{suppl,darko04}.

While the curves $\alpha$ vs $J$ are all qualitatively the same,
there are important features associated with the position of $\mu$,
and thus the value of $\left\langle \rho_{c}\left(0\right)\right\rangle $.
Specifically for $\mu=-2.0t$ we enter the NFL region for relatively
high values of the Kondo coupling, $J\simeq2.35t$, and with an average
$f$-level occupation $\left\langle n_{f}\right\rangle \simeq0.89$
not so close to unity (for all the other values of $\mu$ considered
$\left\langle n_{f}\right\rangle \simeq1$). Moreover, for $J=2.2t$
the thermodynamic properties diverge as a power law with an exponent
$1-\alpha\simeq0.4$, but if we then vary $\mu$ by $10\%$ we get
$\alpha\gg1$ and the system displays FL behavior.

To understand how a power-law distribution of Kondo temperatures emerges
in this problem, we closely follow the arguments of Ref. \cite{darko04}.
In the Kondo limit, $\left\langle n_{f}\right\rangle \rightarrow1$
and $J\rightarrow0$, it is easy to show that $\tk^{\ell}=\tk^{0}\mbox{exp}\left[-\theta_{\ell}^{2}\right]$,
where $\theta_{\ell}^{2}=\pi\Delta_{c\ell}^{\prime}\left(0\right)^{2}/J\left\langle \Delta_{c\ell}^{\prime\prime}\left(0\right)\right\rangle $
and $\tk^{0}=D\mbox{exp}\left[-\pi\left\langle \Delta_{c\ell}^{\prime\prime}\left(0\right)\right\rangle /J\right]$
\cite{suppl}. Here $D$ is an energy cutoff and $\Delta_{c\ell}\left(\omega\right)\equiv\omega-1/G_{\ell\ell}^{c}\left(\omega\right)$
is the local $c$-electron cavity function \cite{vlad98} with a single
(double) prime denoting its real (imaginary) part. For $\Delta_{c\ell}^{\prime}\left(0\right)$
distributed according to a Gaussian (see the inset of Fig.~\ref{fig:alpha}),
it then follows immediately that, up to logarithmic corrections, $P(\tk)\propto\tk^{\alpha-1}$,
with $\alpha=J\left\langle \Delta_{c}^{\prime\prime}\left(0\right)\right\rangle /2\pi\sigma_{c}^{2}$,
where $\sigma_{c}$ is the variance of $P\left(\Delta_{c}^{\prime}\left(0\right)\right)$
\cite{suppl}. Physically, $\Delta_{c\ell}^{\prime}\left(0\right)$
can be interpreted as a renormalized on-site site energy for the $c$ electrons.
The simple Gaussian form of $P\left(\Delta_{c}^{\prime}\left(0\right)\right)$,
as in the usual disordered problem \cite{darko04}, suggests an effective
self-averaging, in the sense that for local quantities like $\Delta_{c}^{\prime}\left(0\right)$
there seems to be no important distinction between disorder and quasiperiodic
order. Nevertheless, we know that this surprising result \emph{cannot}
hold for all observables, since, e.g., transport in quasicrystals
is known to display ``superdiffusive'' behavior \cite{yuan00,grimm03,anu06}.

%%%%%%%%%%%%%%%%%%%%%%%%%%%%%%%%%%%%%%%%%%%%%%%%%%%%%%%%%%%%%%%%%%%%%%%

\emph{Finite-size effects and NFL behavior at finite temperatures.}---
To check the robustness of our scenario against finite-size effects,
we performed simulations on approximants of different sizes $N_{a}$.
For all approximants, we find a minimum Kondo temperature in the sample,
$\tkm$. Below $\tkm$, FL behavior is then restored within our model
(all local moments are screened). From Fig.~\ref{fig:ptk}, we learn
that the power-law distribution of Kondo temperature $P(\tk)\propto\tk^{\alpha-1}$
emerges for $\tk<\tkt$. Taken together, these two observations imply,
in principle, that the NFL range is restricted to the interval $\tkm<T<\tkt$.
However, our calculations show that $\tkm$ vanishes as $N_{a}$ increases
while $\tkt$ remains finite. We thus conclude that the NFL range
actually extends down to $T=0$ in an infinite quasicrystal \cite{suppl}.

\begin{figure}[t]
\begin{centering}
\includegraphics[scale=0.27]{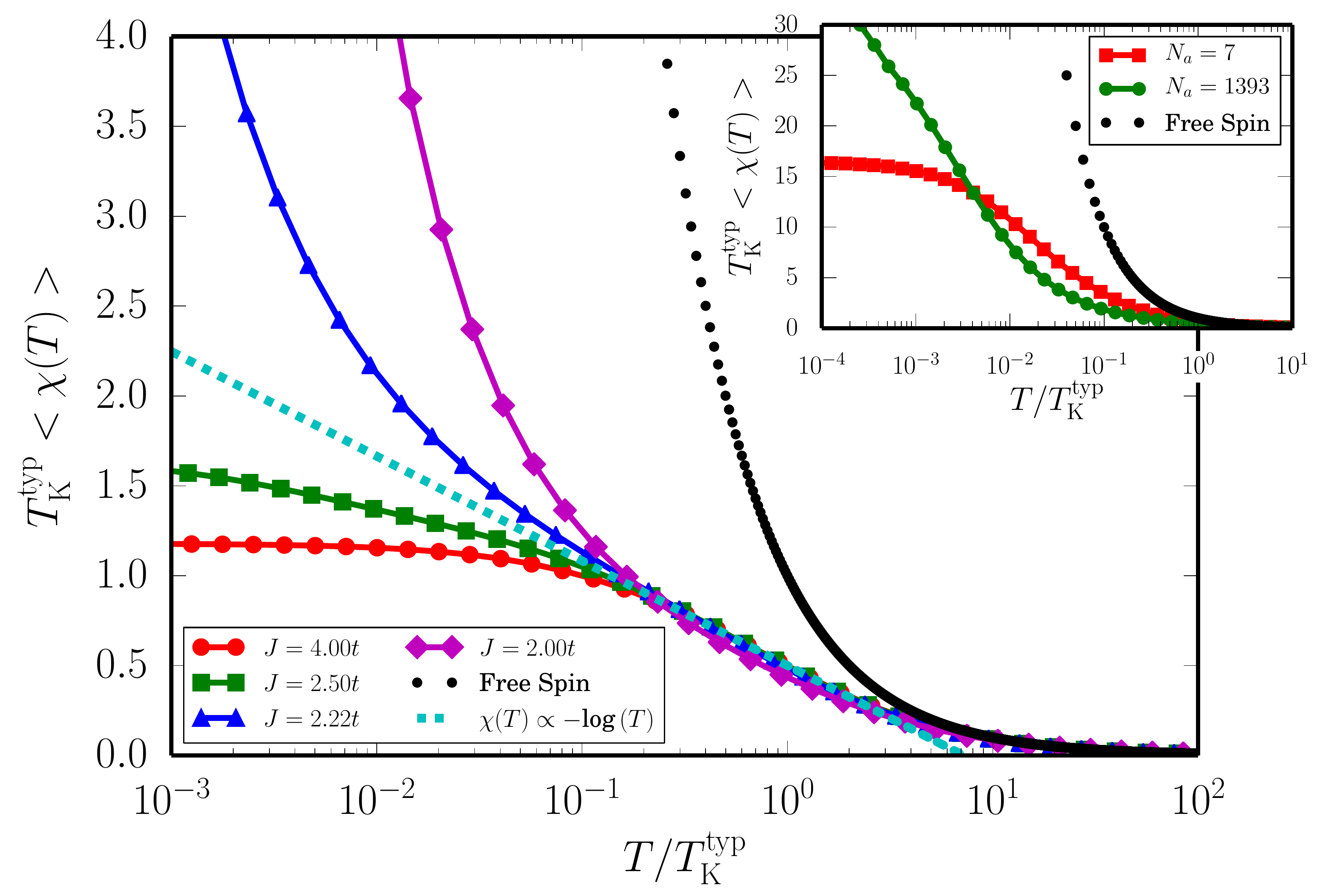} 
\par\end{centering}

\caption{\label{fig:chiT} Averaged value of the impurity susceptibility $\left\langle \chi\left(T\right)\right\rangle $
times the typical value of the Kondo temperature $\tkt$ as a function
of the temperature $T$ normalized by $\tkt$ for four values of the
Kondo coupling $J$ on a semilog scale. For completeness, we show
both the free spin and the $\chi\propto-\mbox{log}\left(T\right)$
$\left(\alpha=1\right)$ curves. Here we considered $\mu=-2.0t$,
$N_{a}=1393$, and $N_{\phi}=576$. Inset: $\tkt\left\langle \chi\left(T\right)\right\rangle $
as a function of $T/\tkt$ at $\mu=2.2t$ and $J=1.05t$ for two different
approximant sizes: $N_{a}=7$ and $N_{a}=1393$. }
\end{figure}

To access the finite-temperature behavior of the system and to observe
the anticipated NFL behavior, we consider a simple interpolative formula
for the local-moment susceptibility, $\chi(T,\tk)=1/(T+\tk)$, which
captures the leading behavior at both low and high-$T$ \cite{Hewson_kondo,suppl}.
We then calculate the magnetic susceptibility of dilute moments as
an average of single-impurity contributions, $\langle\chi(T)\rangle=N_{a}^{-1}\sum_{\ell=1}^{N_{a}}\chi\left(T,\tk^{\ell}\right)$,
with sample results in Fig.~\ref{fig:chiT} \cite{nfl}. In the region
$T\gg\tkt$, $\left\langle \chi\left(T\right)\right\rangle $ shows
the expected free-spin form for all values of the Kondo coupling.
For $T\ll\tkt$, and for $N_{a}\rightarrow\infty$, we observe two
distinct behaviors depending on the value of $\alpha$. For $\alpha>1$
we recover the FL behavior at low-$T$ with $\left\langle \chi\left(T\right)\right\rangle \sim1/\tkt$,
whereas for $\alpha<1$ we obtain $\left\langle \chi\left(T\right)\right\rangle \propto T^{\alpha-1}$.
Moreover, in the crossover region, $T\sim\tkt$, we have the surprisingly
robust result $\left\langle \chi\left(T\right)\right\rangle \sim-\mbox{log}\left(T\right)$,
regardless of the value of $\alpha$. This is due to the fact that
$P(\tk)$ is essentially flat around $\tkt$ (Fig.~\ref{fig:ptk}).
%%%%% Therefore, this range where  reflects the region over which  is basically ``flat''. Therefore, we expect a robust logarithm dependence for , because  is really broad, and as one changes parameters  only the low- range is affected %%%%%
 For the smaller approximants, however, $\tkm$ is finite and hence
FL behavior must be restored at $T<\tkm$ for all $J$. This is explicitly
shown in the inset of Fig.~\ref{fig:chiT} where $\tkm\simeq10^{-2}\tkt$
for $N_{a}=7$.

\emph{Electronic Griffiths phase and }$\aualyb$\emph{.}---The Kondo-disorder(like)
scenario discussed here nicely accounts for power-law divergences
in the thermodynamic quantities when dilute $\mbox{Yb}$ local moments
are placed in a metallic quasicrystal. However, the quasicrystal $\aualyb$
forms a dense Kondo lattice, and one may wonder to what extent our
scenario is relevant in this context. Based on analogies with disordered
Kondo systems (where both the dilute-impurity case and the lattice
case produce $P\left(\tk\right)\propto\tk^{\alpha-1}$ \cite{eduardo96,eduardo01,darko04,kaul07}),
we then expect power-law distributions of Kondo temperatures and the
corresponding NFL phenomenology for $\chi$ and $C/T$ also for the
lattice problem. In that case, the NFL region is known as an electronic
quantum Griffiths phase and it has by now been observed in several
disordered strongly correlated systems \cite{amd09,carol13}.

The quasicrystal heavy fermion $\aualyb$ shows NFL behavior with
$\chi\sim T^{-0.51}$, $C/T\sim-\mbox{log}\left(T\right)$ \cite{deguchi12}
or $\chi\sim T^{-0.55}$, $C/T\sim T^{-0.66}$ \cite{watanuki12}.
Our results, however, predict the same NFL exponent for both $\chi$
and $C/T$, and this difference hampers a definite identification
of quantum Griffiths effects \cite{thomas10}. On the other hand,
the (Griffiths) power-law divergences are exact only at asymptotically
low temperatures, where the regular contribution to the thermodynamic
responses may be completely disregarded, and in general the results
depend not only on the full form of the $P(\tk)$ curve but also on
the particular shape of the scaling functions for the physical observables
\cite{suppl,nfl}, which may account for differences in the exponent.
One such example is the transient $-\mbox{log}\left(T\right)$ divergence
in $\left\langle \chi\left(T\right)\right\rangle $, which is present
for all values of the exponent $\alpha$ in the region $T\sim\tkt$,
Fig. \ref{fig:chiT}. %%%% Another important feature of the  alloy is that the NFL behavior is insensitive to pressure but it is quenched by an external magnetic field . Electronic Griffiths phases have at their origin the existence of rare-regions . Because those are localized and disconnected regions in which  collectively  it is then reasonable to assume that their dynamics is insensitive to pressure (applying pressure does not connect these far away regions), but that it is affect by an external field (it quenches the  entropy of the free spins). [??? TRUE ???] %%%%

Interestingly, it was also reported that the temperature dependence
of $\chi$ and $C/T$ of the quasicrystal $\aualyb$ differs from
that of its crystalline approximant. Reference~\onlinecite{deguchi12}
observes no NFL behavior for the approximant, whereas Ref.~\onlinecite{watanuki12}
does observe NFL behavior but with different powers as compared to
the quasicrystal. To briefly address this intriguing result, we first
notice that the size of the approximant unit cell considered in \cite{deguchi12,watanuki12}
is small and thus it is reasonable to assume that the experimental
situation is similar to the one illustrated in the inset of Fig. \ref{fig:chiT},
where the NFL behavior is bound to be observed only in a relatively
narrow range $\tkm\lesssim T\lesssim\tkt$. Moreover, due to the strong
energy dependence of $\left\langle \rho_{c}\left(\omega\right)\right\rangle $,
Fig. \ref{fig:qc}(c), especially for $\mu$ close to a dip (which
seems to be case for $\aualyb$ \cite{jazbec14}), tiny variations
in parameters, such as the band filling or Kondo coupling, may drive
the system to or from a NFL behavior. Therefore, care should be taken
when drawing any conclusions from this distinct behavior.

\emph{Conclusions.}---Motivated by the recently observed NFL behavior
in the heavy-fermion quasicrystal $\aualyb$, we investigated the
single-impurity Kondo effect in the octagonal $\left(2D\right)$ and
icosahedral $\left(3D\right)$ tilings. We found a power-law distribution of Kondo temperatures $P(\tk)\propto\tk^{\alpha-1}$
and corresponding NFL behavior, in a surprising similarity to disordered
metals. Therefore, a quasicrystalline conduction band provides a natural
route to the emergence of a robust NFL behavior without the tuning
of external parameters as doping, pressure, or external field. For
the Kondo quasicrystalline lattice problem, we expect, based on the
analogy to disordered systems \cite{nfl_review05}, a similar NFL
behavior to be observed. In addition, it would be interesting to investigate
the feedback effect of the local moments, in particular moments with
$\tk<T$, on the transport properties of the quasicrystalline conduction
electrons and the effects of intersite spin correlations \cite{darko05}. 

%%%% However, it is not clear how the formation of rare-regions takes place in a quasiperiodic lattice, due to the existence of only a finite number of local environments %%%

\begin{acknowledgments}
 
We gratefully acknowledge help from M. Mihalkovic
with approximants of the 3D icosahedral tiling. A.J. thanks P. Coleman
for useful discussions on the quasiperiodic heavy fermion $\aualyb$.
E.C.A. was supported by FAPESP (Brazil) Grant No. 2013/00681-8. E.M. was
supported by CNPq (Brazil) Grants No. 304311/2010-3 and No. 590093/2011-8.
M.V. was supported by the Helmholtz association through VI-521 and
by the DFG Grants No. GRK 1621 and No. SFB 1143. V.D. was supported by the
NSF Grants No. DMR-1005751 and No. DMR-1410132.

\end{acknowledgments}

%%%%%%%%%%%%%%%%%%%%%%%%%%%%%%%%%%%%%%%%%%%%%%%%%%%%%%%%%%%%%%%%%%%%%%%

%%%%%%%%%%%%%%%%%%%%%%%%%%%%%%%%%%%%%%%%%%%%%%%%%%%%%%%%%%%%%%%%%%%%%%%


\begin{thebibliography}{61}
\bibitem[{tk()}]{tk}
\bibinfo{note}{Because the $U\rightarrow\infty$ Anderson impurity model is
  particle-hole asymmetric, this definition of $T_{\rm K}$ is, in general,
  close but not equal to the onset temperature of a non-trivial solution
  $\left(Z_{\ell}\neq0,\:\tilde{\varepsilon}_{f\ell}\neq E_{f}\right)$ of the
  saddle-point equations.}
\bibitem[{nfl()}]{nfl}
\bibinfo{note}{Using this interpolative formula it is easy to see that
  $\left\langle \chi\left(T\right)\right\rangle =\int dT_{\rm K}P\left(T_{\rm
  K}\right)/\left(T+T_{\rm K}\right)=\chi_{r}+\chi_{s}$. $\chi_{r}$ is a
  regular contribution whereas $\chi_{s}\propto T^{\alpha-1}$ gives raise to
  NFL behavior for $\alpha<1$ as $T\rightarrow0$.}
\end{thebibliography}
\end{document}

\title{Supplementary information for:\\
 ``Non-Fermi-liquid behavior in metallic quasicrystals with local
magnetic moments''}

\author{Eric C. Andrade}

\affiliation{Instituto de F\'{i}sica Te\'{o}rica, Universidade Estadual Paulista, Rua
Dr. Bento Teobaldo Ferraz, 271 - Bl. II, 01140-070, S\~{a}o Paulo, SP,
Brazil}

\author{Anuradha Jagannathan}

\affiliation{Laboratoire de Physique des Solides, CNRS-UMR 8502, Universit\'{e} Paris-Sud,
91405 Orsay, France}

\author{Eduardo Miranda}

\affiliation{Instituto de F\'{\i}sica Gleb Wataghin, Unicamp, Rua S\'{e}rgio Buarque de Holanda, 777, CEP 13083-859 Campinas, SP, Brazil}

\author{Matthias Vojta}

\affiliation{Institut f\"{u}r Theoretische Physik, Technische Universit\"{a}t Dresden,
01062 Dresden, Germany}

\author{Vladimir Dobrosavljevi\'{c}}

\affiliation{Department of Physics and National High Magnetic Field Laboratory,
Florida State University, Tallahassee, FL 32306}

%%%%%%%%%%%%%%%%%%%%%%%%%%%%%%%%%%%%%%%%%%%%%%%%%%%%%%%%%%%%%%%%%%%%%%%

\date{\today}

\maketitle
%%%%%%%%%%%%%%%%%%%%%%%%%%%%%%%%%%%%%%%%%%%%%%%%%%%%%%%%%%%%%%%%%%%%%%%

 %%% Here we discuss extra properties of the octagonal tiling, further details of the numerical simulation, and the effective theory for the Griffiths phase. %%%

%%%%%%%%%%%%%%%%%%%%%%%%%%%%%%%%%%%%%%%%%%%%%%%%%%%%%%%%%%%%%%%%%%%%%%%

\section{Electronic wavefunctions in the octagonal tiling}

Based on the results for one-dimensional quasicrystals,\cite{kohmoto87}
we expect the resulting wavefunctions in quasiperiodic tight-binding
models to be different both from the exponentially localized wavefunctions 
found in Anderson insulators as well from Bloch states found
in a crystal. Such wavefunctions are the so-called \emph{critical}
wavefunctions. In real space, this means very large fluctuations
of the wavefunction amplitude from site to site but with similar
amplitudes on sites of similar local environment (the amplitude distribution
is thus determined by the deterministic scale invariant geometry).
%%% with an average behavior with decay slowly (as a power-law) in real space%%%

To probe the real space profile of the wavefunctions, we compute
the inverse participation ratio %%% ,or the second moment of the probability density %%%
\begin{eqnarray}
P_{\nu}^{-1} & = & \sum_{i}\left|\psi_{\nu}^{c}\left(i\right)\right|^{4},\label{eq:ipr}
\end{eqnarray}
where $\psi_{\nu}^{c}$ is an eigenstate of $\mathcal{H}_{c}$ (defined
in Eq. {[}1{]} of the main text) with energy $E_{\nu}^{c}$. The scaling
of $P_{\nu}^{-1}$ with the system size is related to the spatial
structure of the electronic states. If we write $P_{\nu}^{-1}\propto N_{a}^{-\beta}$,
then $\beta=1$ for extended and $\beta=0$ for exponentially localized
states. In a quasicrystal, because of the critical nature of the 
wavefunctions, we expect that $0\le\beta\le1$. It is important to point
out that the converse it is not true. For instance, $\beta\approx1$
does not necessarily imply an extended state. To establish such result,
we need to study the scaling of the higher moments of the wavefunction
distribution because the exponents that describe the scaling of these
moments with $N_{a}$ are not just multiples of each other \textendash{}
a property of multifractality.\cite{chhabra89,richardella10,rodriguez10}
As mentioned above, this resulting multifractal character of 
wavefunctions is a consequence of the invariance of quasicrystals under
scale transformations, a feature called inflation-deflation symmetry.\cite{anu06}

In Fig. \ref{eq:ipr} we calculate $P_{\nu}^{-1}$ at different positions
in the band (but away form the band center) and obtain $\beta\approx0.90$,
which is close but smaller than one and consistent with a preponderance
of multifractal eigenstates. Remarkably, this is similar to the value
of $\beta$ reported for the Penrose tiling.\cite{grimm03}

\begin{figure}[t]
\begin{centering}
\includegraphics[scale=0.28]{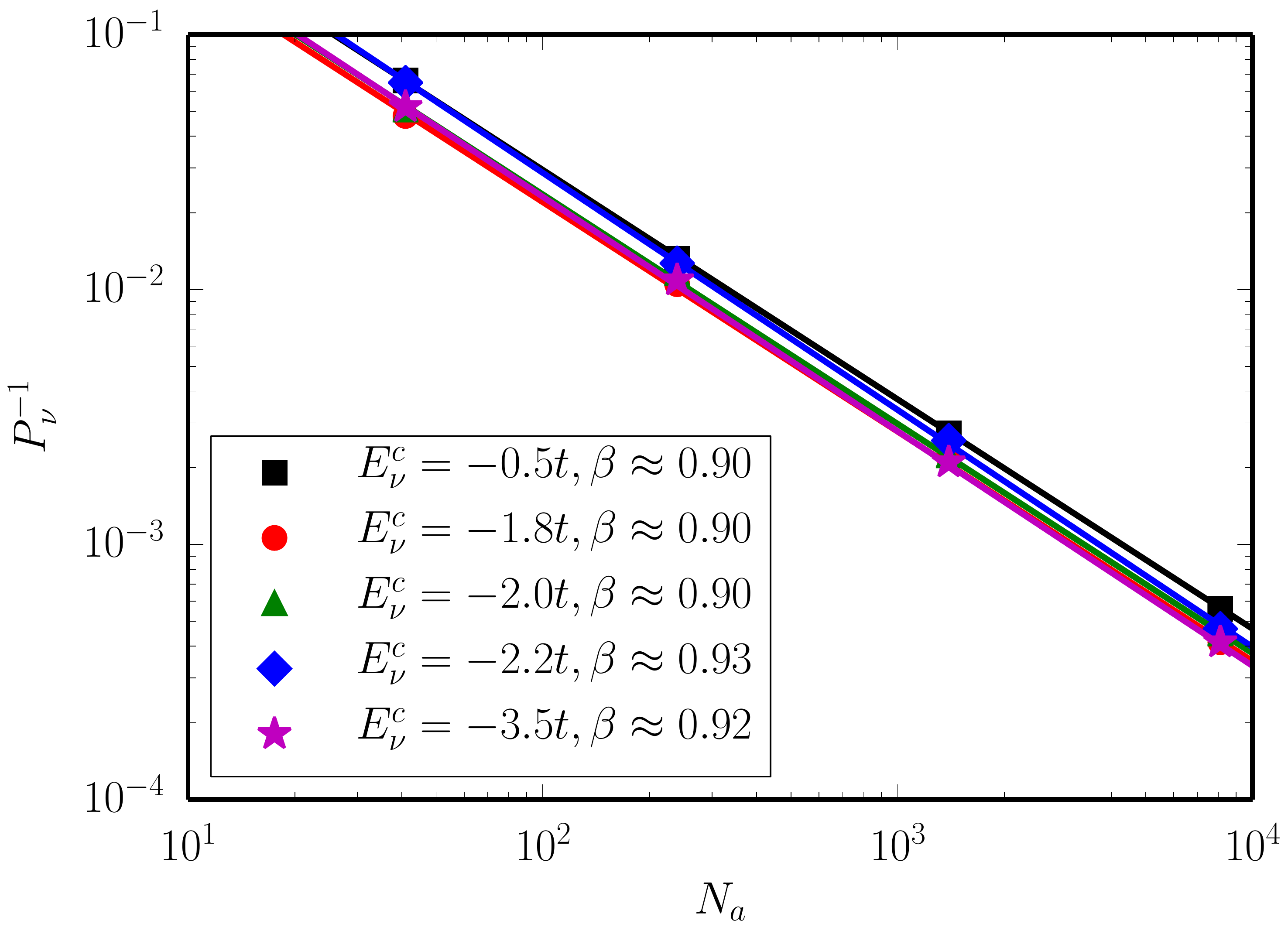}
\par\end{centering}

\caption{\label{fig:ipr} Octagonal tiling. Inverse participation ratio $P_{\nu}^{-1}$
as a function of the approximant size $N_{a}$ for different values
of the eigenenergies $E_{\nu}^{c}$ on a log-log scale. We considered
four approximant sizes $N_{a}=41$, $239$, $1393$, $8119$. The
solid lines are power-law fits: $P_{\nu}^{-1}\propto N_{a}^{-\beta}$.}
\end{figure}

\begin{figure}[t]
\begin{centering}
\includegraphics[scale=0.28]{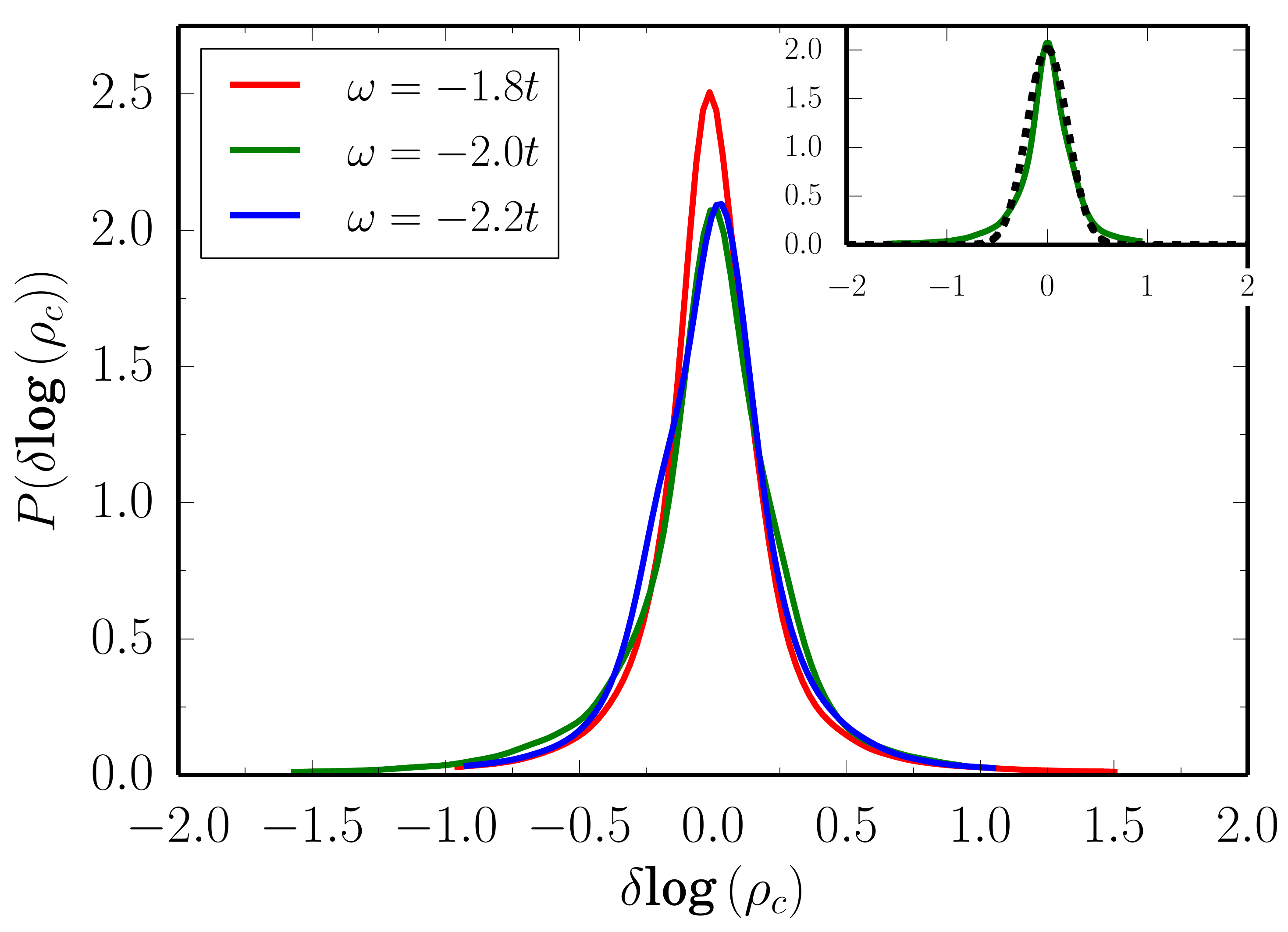}
\par\end{centering}

\caption{\label{fig:P_logrho} Octagonal tiling. Distribution of $\delta\mbox{log}\left(\rho_{c}\right)=\mbox{log}\left(\rho_{c}\right)-\left\langle \mbox{log}\left(\rho_{c}\right)\right\rangle $
at three different values of the energy $\omega$. Inset: The full
curve correspond to $P\left(\delta\mbox{log}\left(\rho_{c}\right)\right)$
at $\omega=-2.0t$ and the dashed curve is a Gaussian fit to it. Here
we considered the $N_{a}=8119$ approximant with $N_{\phi}=64$. }
\end{figure}

Another useful quantity that probes the nature of the wavefunction
is the local density of states (LDOS)
\begin{equation}
\rho_{i}^{c}\left(\omega\right)=\overline{\sum_{\nu}\left|\psi_{\nu}^{c}\left(i\right)\right|^{2}\delta\left(\omega-E_{\nu}^{c}\right)},\label{eq:ldos}
\end{equation}
where the overline denotes average over boundary conditions. The distribution
of the logarithm of $\rho_{i}^{c}\left(\omega\right)$ is presented
in Fig. \ref{fig:P_logrho}. As we can see, the curves are all qualitatively
the same, indicating that the spatial fluctuations of $\rho_{i}^{c}$
are, to a good extent, energy-independent and well described by a
log-normal distribution (see the inset). Specifically, the width of
the distribution does not vary much with the energy, except for $\omega=-2.0t$
where there is a slightly larger tendency to have $\rho_{i}^{c}\left(\omega\right)$
smaller than its typical value.

Generally, a log-normal distribution of LDOS is expected to occur
in an Anderson insulator,\cite{mirlin00} but it is also known that
a log-normal also nicely describes the distribution of LDOS of disordered
metals.\cite{schubert10} Therefore, a careful finite-size scaling
study is required to establish the precise nature of the wavefunction
based on $P\left(\rho\right)$.\cite{schubert10} We leave this more
detailed investigation for a future work.

To conclude the discussion on the octagonal tiling, we briefly mention
our implementation of averaging over twisted boundary conditions (TBC).\cite{gross96,assad02}
This step was instrumental in obtaining reliable results since it
provides a controlled way to eliminate finite-size effects associated
with spectral discreteness (we will come back to the role of finite-size
effects later). The average over TBC is completely equivalent to a
periodic repetition of the $N_{a}$-site approximant in both directions
using $N_{\phi}$ copies, hence the effective total linear system
size is increased to $\sqrt{N_{a}N_{\phi}}$.

%%%%%%%%%%%%%%%%%%%%%%%%%%%%%%%%%%%%%%%%%%%%%%%%%%%%%%%%%%%%%%%%%%%%%%%

\section{Slave bosons mean-field equations and the Kondo limit}

Minimizing the slave boson (SB) mean-field free energy, Eq. {[}2{]}
in the main text, with respect to $\tilde{\varepsilon}_{\ell}$ and
$Z_{\ell}$ we obtain the corresponding self-consistency equations

\begin{eqnarray}
\frac{2}{\pi}\int_{-\infty}^{+\infty}f\left(\omega\right)\mbox{Im}\left[\tilde{G}_{\ell}^{f}\left(\omega\right)\right]d\omega+Z_{\ell}-1 & = & 0,\label{eq:MF1-a}\\
\frac{2}{\pi}\int_{-\infty}^{+\infty}f\left(\omega\right)\mbox{Im}\left[\tilde{G}_{\ell}^{f}\left(\omega\right)\Delta_{f\ell}\left(\omega\right)\right]d\omega & +\nonumber \\
\tilde{\varepsilon}_{f\ell}-E_{f} & = & 0.\label{eq:MF2-a}
\end{eqnarray}
In general, we solve these mean-field equations numerically at $T=0$.
They are algebraic non-linear equations on the parameters $Z_{\ell}$
and $\tilde{\varepsilon}_{f\ell}$, which we solve using a globally
convergent implementation of the Newton-Raphson algorithm.\cite{numerical_recipes}
The integral over frequencies is performed using the Romberg method.\cite{numerical_recipes}

If we now move to the Kondo limit, where both $Z_{\ell},\,\tilde{\varepsilon}_{f\ell}\rightarrow0$,
we are able to solve Eqs. \eqref{eq:MF1-a} and \eqref{eq:MF2-a}
analytically. In this limit, we ignore the frequency dependence of
the hybridization function, $\Delta_{f\ell}\left(\omega\right)\simeq\Delta_{f\ell}^{\prime}\left(0\right)+i\Delta_{f\ell}^{\prime\prime}\left(0\right)$,
and assume that the integrals are dominated by their values at the
Fermi level. From Eq. \eqref{eq:MF1-a} we obtain
\begin{eqnarray}
\tilde{\varepsilon}_{f}+Z_{\ell}\Delta_{f\ell}^{\prime}\left(0\right) & \simeq & 0.\label{eq:tildeps_kondo}
\end{eqnarray}
which reflects the well-known fact that in the Kondo limit the position
of the Kondo peak, $\tilde{\varepsilon}_{f}+Z_{\ell}\Delta_{f\ell}^{\prime}\left(0\right)$,
moves to the Fermi level.

From Eq. \eqref{eq:MF2-a} we can now calculate the Kondo temperature
(recall our definition $\tk^{\ell}\equiv Z_{\ell}\Delta_{f\ell}^{\prime\prime}\left(0\right)$)
\begin{eqnarray}
\tk^{\ell} & = & D\mbox{exp}\left[-\frac{\pi}{2}\frac{\Delta_{f\ell}^{\prime}\left(0\right)+\left|E_{f}\right|}{\Delta_{f\ell}^{\prime\prime}\left(0\right)}\right],\label{eq:tk_kondo}
\end{eqnarray}
where $D$ is an high-energy cutoff of the integral of the order of
the bandwidth. We recover the usual Kondo expression, $\tk^{\ell}=D\mbox{exp}\left[-1/J\rho_{\ell}^{c}\left(0\right)\right]$,
in the case of particle-hole symmetry, $\Delta_{f\ell}^{\prime}\left(0\right)=0$,
with $\Delta_{f\ell}^{\prime\prime}\left(0\right)=\pi V^{2}\rho_{\ell}^{c}\left(0\right)$
and $J=2V^{2}/\left|E_{f}\right|$.

%%%%%%%%%%%%%%%%%%%%%%%%%%%%%%%%%%%%%%%%%%%%%%%%%%%%%%%%%%%%%%%%%%%%%%%

\section{Asymptotic expression for $P\left(\tk\right)$}

Each site of the tiling has a different local $c$-electron cavity
function $\Delta_{c\ell}\left(\omega\right)$, reflecting the fact
that the effective potential that one electron sees as it goes through
the lattice changes from site to site. If we go one step further,
we may consider its real part at the Fermi level, $\Delta_{c\ell}^{\prime}\left(0\right)$,
as a renormalized on-site site energy for the $c$-electrons. According
to the arguments presented in Ref.~\onlinecite{darko04} for the
case of weakly disordered Kondo systems, a power-law distribution
for the Kondo temperature can be easily obtained provided that the
fluctuations of $\Delta_{c\ell}^{\prime}\left(0\right)$, $\delta\Delta_{c}^{\prime}\equiv\Delta_{c}^{\prime}\left(0\right)-\left\langle \Delta_{c}^{\prime}\left(0\right)\right\rangle $,
follow a Gaussian distribution (see the inset of Fig. 3 of the main
text). In disordered systems, the fluctuations of the local $c$-electron
cavity function at a given site $i$ result from Friedel oscillations
of the electronic wavefunctions induced by other impurities which
may lie at a relatively long distance from $i$. Furthermore, at weak
disorder, $\delta\Delta_{c\ell}^{\prime}$ takes the form of a linear
superposition of contributions from single impurity scatterers, and
thus of a sum of independent random numbers, for which we expect the
central limit theorem to hold. From our numerical results, we then
reason that a similar mechanism takes place in quasicrystals. This
somewhat surprising resemblance between a quasicrystal and weakly
disordered systems, rather than systems at the metal-insulator transition,
is also present in different physical quantities, e.g. the level-spacing
distribution.\cite{grimm03,anu06,anu00} It indicates that a quasicrystal
in higher dimensions may show a more conventional behavior in local
quantities despite its multifractal eigenstates.

It is now a straightforward exercise to obtain the asymptotic expression
$P\left(\tk\right)\sim\tk^{\alpha-1}$ following Ref.~\onlinecite{darko04}.
We start by relating the $f-$level hybridization function $\Delta_{f\ell}$
with the local $c$-electron cavity function $\Delta_{c\ell}$ at
the impurity site $\ell$
\begin{eqnarray}
\Delta_{f\ell}^{\prime}\left(\omega\right) & = & \frac{V^{2}\left(\omega-\Delta_{c\ell}^{\prime}\left(\omega\right)\right)}{\left(\omega-\Delta_{c\ell}^{\prime}\left(\omega\right)\right)^{2}+\left(\Delta_{c\ell}^{\prime\prime}\left(\omega\right)\right)^{2}},\label{eq:re_deltaf}\\
\Delta_{f\ell}^{\prime\prime}\left(\omega\right) & = & \frac{V^{2}\Delta_{c\ell}^{\prime\prime}\left(\omega\right)}{\left(\omega-\Delta_{c\ell}^{\prime}\left(\omega\right)\right)^{2}+\left(\Delta_{c\ell}^{\prime\prime}\left(\omega\right)\right)^{2}},\label{eq:im_deltaf}
\end{eqnarray}
where, as usual, single (double) primes denote the real (imaginary)
part. The next step is to take the Kondo limit making use of Eq. \eqref{eq:tk_kondo}.
As the last assumption, we disregard fluctuation in the imaginary
part of $\Delta_{c\ell}\left(0\right)$, so we replace $\Delta_{c\ell}^{\prime\prime}\left(0\right)$
by its average value $\left\langle \Delta_{c}^{\prime\prime}\left(0\right)\right\rangle $.
Using Eqs. \eqref{eq:re_deltaf} and \eqref{eq:im_deltaf} we then
obtain

\begin{eqnarray}
\tk^{\ell} & = & \tk^{0}\mbox{exp}\left[-\pi\frac{\left(\Delta_{c\ell}^{\prime}\left(0\right)\right)^{2}}{J\left\langle \Delta_{c}^{\prime\prime}\left(0\right)\right\rangle }\right],\label{eq:tk_kondo_1}
\end{eqnarray}
where $\tk^{0}\equiv D\mbox{exp}\left[-\pi\left\langle \Delta_{c}^{\prime\prime}\left(0\right)\right\rangle /J\right]$.
Inverting Eq. \eqref{eq:tk_kondo_1} we may write
\begin{eqnarray}
\delta\Delta_{c\ell}^{\prime} & \simeq & \mbox{ln}^{1/2}\left[\frac{\tk^{0}}{\tk^{\ell}}\right]^{\lambda},\label{eq:delta_cavity}
\end{eqnarray}
with $\lambda=J\left\langle \Delta_{c}^{\prime\prime}\left(0\right)\right\rangle /\pi$
and we also considered the fact that for $\tk^{\ell}\ll\tk^{0}$ we
may drop the the term $\left\langle \Delta_{c\ell}^{\prime}\left(0\right)\right\rangle $.
Since we assume that $P\left(\delta\Delta_{c}^{\prime}\right)$ is
a simple Gaussian with variance $\sigma_{c}$, a direct change of
variables gives, up to a negligible logarithmic correction,
\begin{eqnarray}
P\left(T_{K}\right) & \propto\tk^{\alpha-1} & ,\label{eq:p_tk_qc}
\end{eqnarray}
with
\begin{eqnarray}
\alpha & = & \frac{J\left\langle \Delta_{c}^{\prime\prime}\left(0\right)\right\rangle }{2\pi\sigma_{c}^{2}}\sim J\left\langle \rho_{c}\left(0\right)\right\rangle .\label{eq:alpha}
\end{eqnarray}
So we see that $\alpha$ varies linearly with $J$ with a slope proportional
to $\left\langle \rho_{c}\left(0\right)\right\rangle $.

\begin{figure}[t]
\begin{centering}
\includegraphics[scale=0.28]{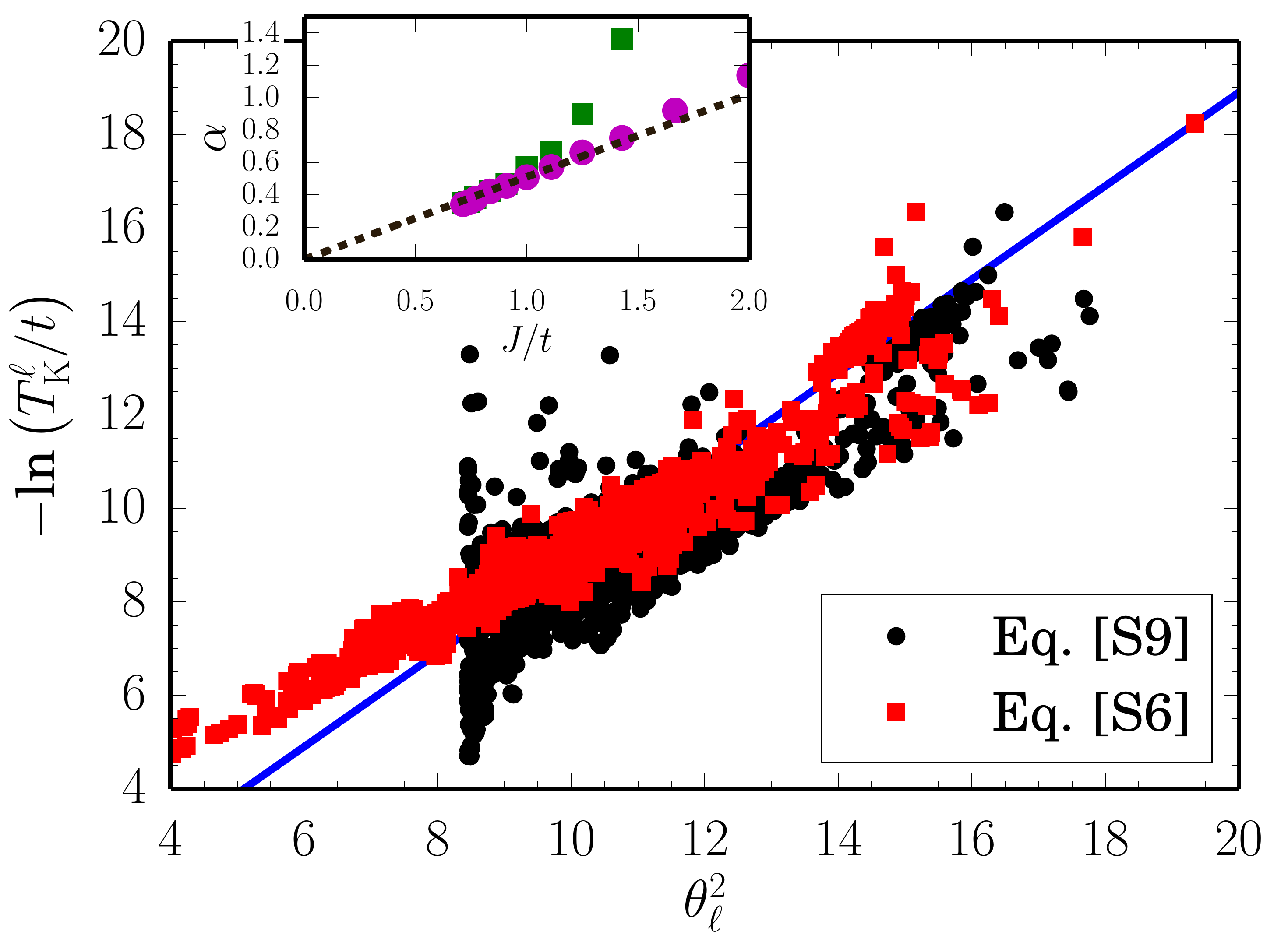}
\par\end{centering}

\caption{\label{fig:tk_scatter} Octagonal tiling. Scatter plot of the numerically
calculated $-\mbox{ln}\left(\tk^{\ell}\right)$ as a function of the
exponent $\theta_{\ell}^{2}$ with $\theta_{\ell}^{2}=\pi\left(\Delta_{f\ell}^{\prime}\left(0\right)+\left|E_{f}\right|\right)/2\Delta_{f\ell}^{\prime\prime}\left(0\right)$
(Eq. \eqref{eq:tk_kondo}) or $\theta_{\ell}^{2}=\pi\left(\Delta_{c\ell}^{\prime}\left(0\right)\right)^{2}/J\left\langle \Delta_{c}^{\prime\prime}\left(0\right)\right\rangle $
(Eq. \eqref{eq:tk_kondo_1}). Each point correspond to a given site
in the approximant and the straight line has a unity slope. Here,
we considered the $N_{a}=1393$ approximant, $J=0.77t$, and $\mu=-2.2t$.
Inset: Power-law exponent $\alpha$ for $\mu=-2.2t$ as a function
of the Kondo coupling $J$. The squares correspond to the exponent
extracted from the numerical data as in Fig. 3 of the main text. The
circles correspond to the exponent extracted from a distribution of
$\tk$ generated according to Eq. \eqref{eq:tk_kondo}. The dashed
line is the asymptotic expression for $\alpha$ in Eq. \eqref{eq:alpha}.
The error bars for $\alpha$ are smaller than the symbol sizes.}
\end{figure}

To check the plausibility of our assumptions, we produced a scatter
plot of the numerically calculated $\tk^{\ell}$ versus the exponent
of the Kondo limit formulas for $\tk^{\ell}$ in Eqs. \eqref{eq:tk_kondo}
and \eqref{eq:tk_kondo_1}, Fig. \ref{fig:tk_scatter}. There, we
see that all the points (one for each site in the quasicrystal approximant)
follow a straight line, specially as $\tk^{\ell}$ decreases, clearly
indicating a strong correlation between the full numerics and the
asymptotic expressions for the $\tk^{\ell}$. Additional scatter around
this straight line simply reflects departures from Eqs. \eqref{eq:tk_kondo}
and \eqref{eq:tk_kondo_1}, i.e. a situation where the value $\tk^{\ell}$
depends not only on $\Delta_{f\ell}$ at the Fermi level, but on the
entire spectral function. Moreover, as $\tk^{\ell}$ decreases the
curves obtained from Eqs. \eqref{eq:tk_kondo} and \eqref{eq:tk_kondo_1}
become more and more similar showing that the fluctuations in $\Delta_{c\ell}^{\prime}\left(0\right)$
are indeed the dominant ones.

The power-law distribution of Kondo temperatures describes only the
low-$\tk$ tail of the full distribution $P\left(\tk\right)$ and
we then expect that our asymptotic expressions in Eqs. \eqref{eq:p_tk_qc}
and \eqref{eq:alpha} to work better and better as $\alpha$ (or $J$)
diminishes. To check this, we compare the exponent $\alpha$ from
our numerical data with: (i) the exponent extracted from a distribution
of $\tk$ generated according to Eq. \eqref{eq:tk_kondo}; and (ii)
the asymptotic expression for $\alpha$ in Eq. \eqref{eq:alpha}.
In the inset of Fig. \ref{fig:tk_scatter} we show that all three
values of $\alpha$ nicely match for $J\lesssim1t$.

\begin{figure}[t]
\begin{centering}
\includegraphics[scale=0.28]{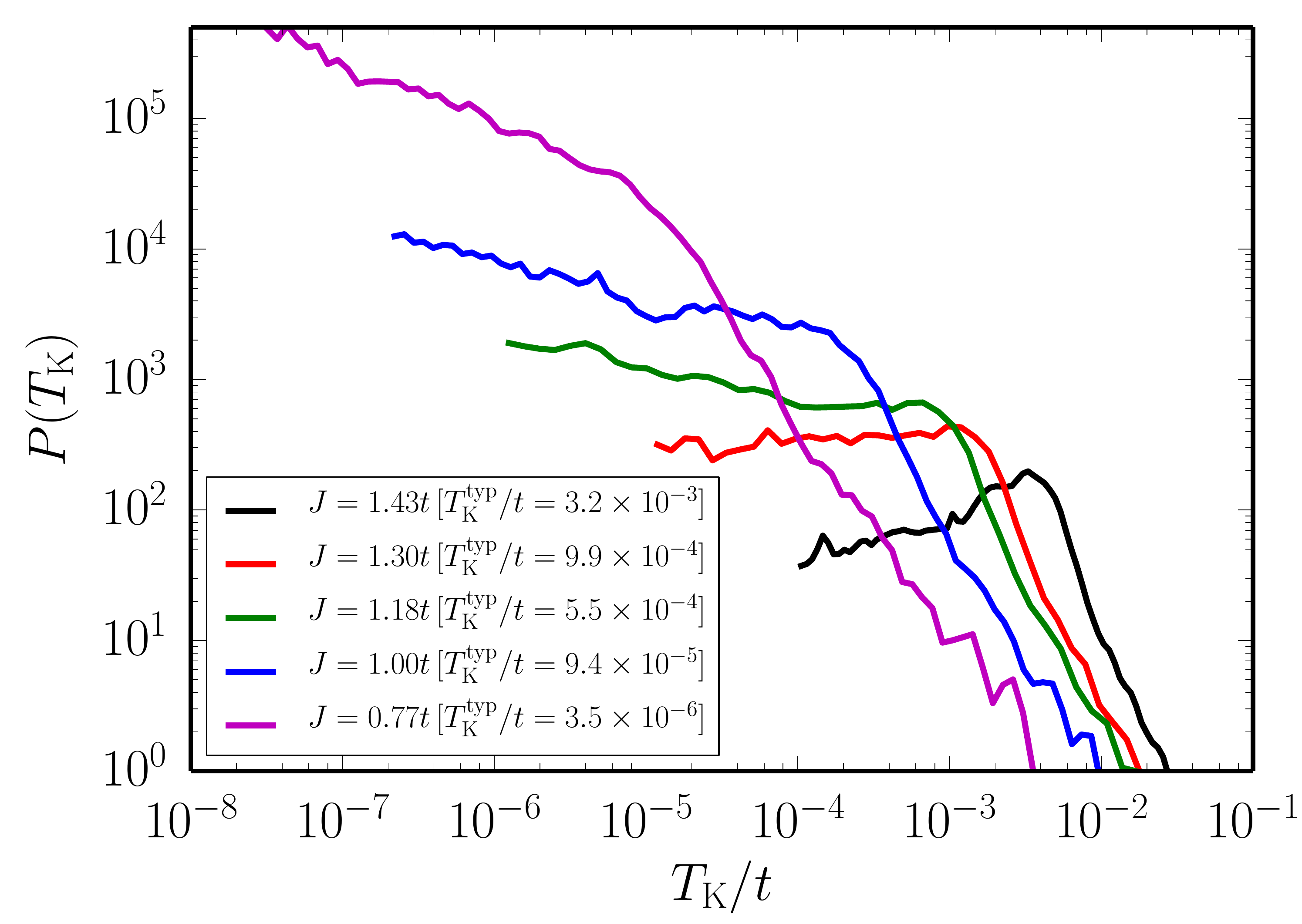}
\par\end{centering}

\caption{\label{fig:ptk} Octagonal tiling. Distribution of the local Kondo
temperatures $P\left(\tk\right)$ as a function of $\tk$ on a log-log
scale for several values of the Kondo coupling $J$. $J$ increases
from the top to the bottom curve. We see that for $\tk\lesssim\tkt$
this distribution acquires a power-law form $P\left(\tk\right)\sim\tk^{\alpha-1}$.
The power-law exponent $\alpha$ continuously varies with the coupling
$J$ and for $\alpha<1$ we have a singular distribution (notice that
for $\tk\gtrsim\tkt$, $P\left(\tk\right)$ is also power-law like,
with a power that does not depend on $J$. This is \emph{not} the
power-law regime we refer to in this work). Here we considered $N_{a}=1393$,
$\mu=-2.2t$, and $N_{\phi}=576$. }
\end{figure}

%%%%%%%%%%%%%%%%%%%%%%%%%%%%%%%%%%%%%%%%%%%%%%%%%%%%%%%%%%%%%%%%%%%%%%%

\section{Singular Kondo temperature distribution and NFL behavior}

As we discussed in the main text, the region in which $\alpha<1$
corresponds to NFL behavior at low-$T$. %%%% Now we discuss this point in more detail and show how a power-law distribution of Kondo temperatures manifests itself in the physical observables at finite-. All our discussion so far has been restricted to the  solution of Eq. . To access finite temperatures, we could easily solve Eq.  at  and calculate the corresponding observables. However, we know that the SB solutions at finite- is problematic, since it treats the onset of the Kondo effect as a phase transition instead of a crossover and assumes that the  and  electrons are decoupled above  . Here instead we follow a complementary route. %%%%%%
To establish this link, we combine our $T=0$ solution of the mean
field equations \eqref{eq:MF1-a} and \eqref{eq:MF2-a} with the well-known
scaling relations for the Kondo impurity problem.\cite{Hewson_kondo}
Essentially, we use the fact that the Kondo problem has a single energy
scale, the Kondo temperature $\tk$, and that the observables can
be written as universal functions of $T/\tk$.

For instance, for the local-moment susceptibility we have
\begin{eqnarray}
\chi\left(T,\tk\right) & \propto & \frac{1}{\tk}f\left(\frac{T}{\tk}\right),\label{eq:chi_scaling}
\end{eqnarray}
with the asymptotic forms of $f\left(x\right)$ given by\cite{Hewson_kondo}
\begin{eqnarray}
f\left(x\right) & = & \left\{ \begin{array}{c}
a-bx^{2}\hspace{2.3cm}x\ll1\\
\left(c/x\right)\left(1-1/\mbox{ln}x\right)\hspace{0.75cm}x\gg1
\end{array}\right.,\label{eq:scaling_f}
\end{eqnarray}
where $a$, $b$, and $c$ are universal numbers. The average value
of the susceptibility is then given by
\begin{eqnarray}
\left\langle \chi\left(T\right)\right\rangle  & = & \int d\tk P\left(\tk\right)\chi\left(T,\tk\right)\nonumber \\
 & = & \chi_{r}+\underbrace{\int_{0}^{\tkma}d\tk\tk^{\alpha-1}\frac{1}{\tk}f\left(\frac{T}{\tk}\right)}_{\propto T^{\alpha-1}}.\label{eq:av_chi}
\end{eqnarray}
Here, $\tkma\sim\tkt$ is a cutoff below which the power-law for of
$P\left(\tk\right)$ holds and we see that $\left\langle \chi\left(T\right)\right\rangle $
contains a regular part $\chi_{r}$ and a potentially singular contribution
$\chi_{s}\propto T^{\alpha-1}$. For $\alpha<1$ and at low-$T$,
we may then disregard $\chi_{r}$ to obtain the anticipated NFL power-law
divergence $\left\langle \chi\left(T\right)\right\rangle \propto T^{\alpha-1}$.
The impurity specific heat divided by the temperature has a similar
behavior and, accordingly, we get $\left\langle C/T\right\rangle \propto T^{\alpha-1}$.
%%% such that the Wilson ratio remains constant.%%%

Given that the SB mean-field approach can be applied at finite-$T$
(albeit resulting in an unphysical finite-temperature transition)
it is then natural to ask ourselves whether it is legitimate to calculate
$P(\tk)$ at $T=0$, and follow the procedure described above, rather
than solving the SB equations at finite-$T$ to explicitly calculate
$\chi\left(T\right)$ and $\gamma\left(T\right)$. %%%% In principle, one could be concerned about the validity of the current procedure, for example if the answer would significantly change if we solved the SB equations at finite T %%%%
From our experience, the general conclusion is that the leading low-$T$
power-law behavior of $\chi$ or $C/T$ is not affected by these additional
effects. Higher-$T$ behavior will of course be affected but as long
as we are interested in leading low-$T$ asymptotics (the value of
the power), the current procedure is well-defined, simply because
the distributions $P\left(\tk\right)$ are very broad. Similar questions
have been raised in the more general context of Quantum Griffiths
Phases and the Infinite-Randomness Fixed Point Behavior.\cite{fisher95}
There again one arrives at a similar conclusion: the $T=0$ distribution
of energy dominates even finite-$T$ behavior.\cite{nfl_review05,thomas06,thomas10}

In the same spirit, we may extend the above discussion to also calculate
observables other than thermodynamical. An interesting quantity to
look at is the the nuclear spin-lattice relaxation rate divided by
temperature $1/\left(T_{1}T\right)$. In our Griffiths scenario, we
expect that $1/\left(T_{1}T\right)\sim T^{\ensuremath{\alpha-2}}\sim\chi/T$.\cite{eduardo96,eduardo97,thomas10}
Nevertheless, the experiment finds that $1/\left(T_{1}T\right)\sim\chi$.\cite{deguchi12}
We point out, however, that this discrepancy is not, at this point,
particularly conclusive, since the curves for $1/\left(T_{1}T\right)$
in Ref. ~\onlinecite{deguchi12} were obtained only for $T>1\, K$,
whereas the NFL behavior is more pronounced for $T<1\, K$. It would
be nice to see how $1/\left(T_{1}T\right)$ behaves at low temperatures,
where it will most certainly provide more conclusive hints as for
the nature of quasicrystalline electronic environment. Another interesting
quantity to investigate is the resistivity.\cite{eduardo96,eduardo97}
However, unlike thermodynamic responses, for which we expect the single
impurity behavior to survive in the dense lattice limit (as is the
case of $\aualyb$), we know that for transport the situation will
be different as coherence between the impurities emerges and thus
we cannot, at this point, compare our predictions to the experiments.
Furthermore, we expect this to be a non-trivial problem, because even
in the absence of correlations, transport in quasicrystals is known
to display an unusual ``super-diffusive'' behavior.\cite{yuan00,grimm03,anu06}

%%%%%%%%%%%%%%%%%%%%%%%%%%%%%%%%%%%%%%%%%%%%%%%%%%%%%%%%%%%%%%%%%%%%%%%

\section{Size dependence of $P\left(\tk\right)$}

\begin{figure}[t]
\begin{centering}
\includegraphics[scale=0.28]{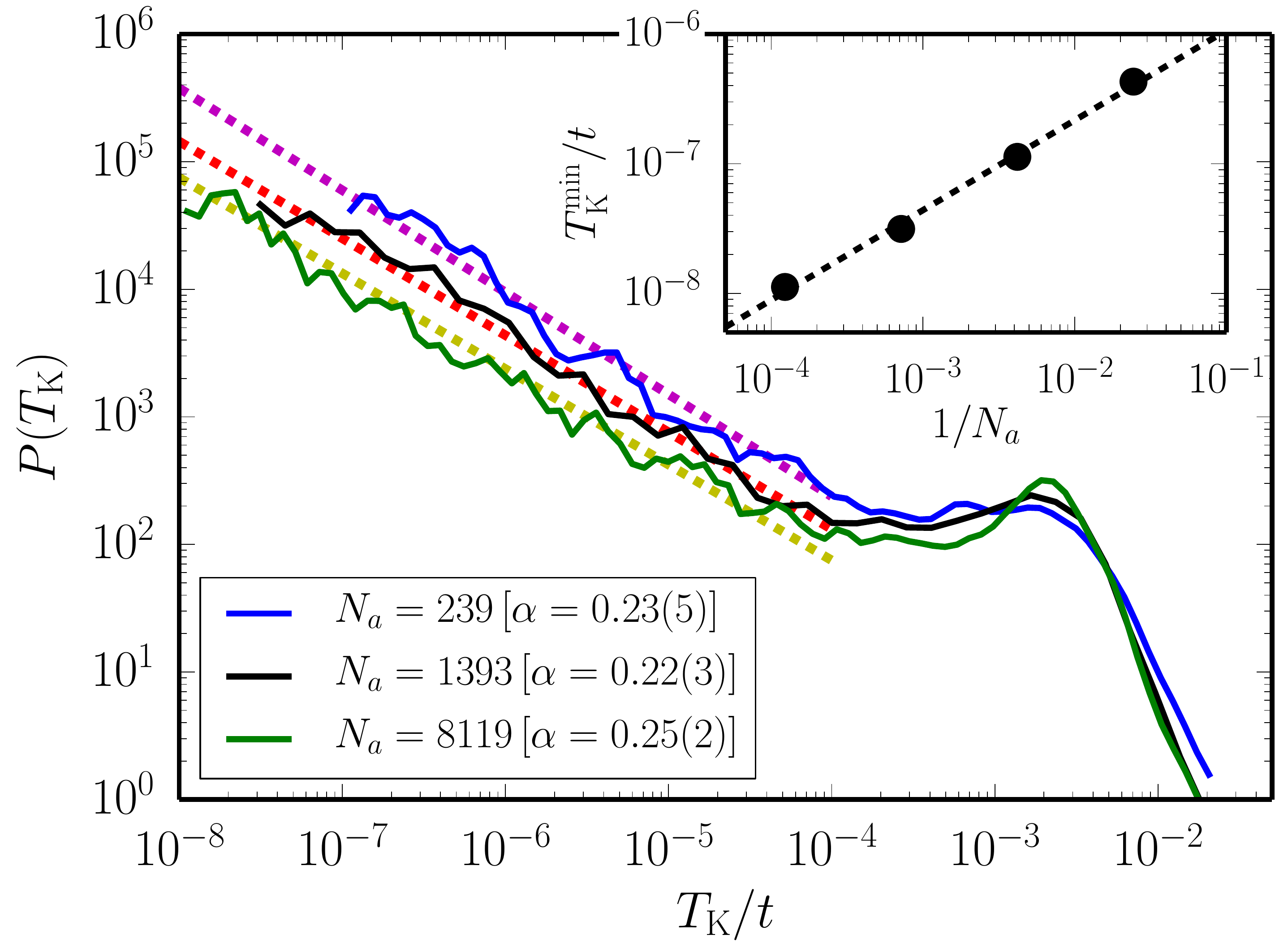}
\par\end{centering}

\caption{\label{fig:PTK_versusN} Octagonal tiling. Distribution of Kondo temperatures
$P\left(\tkma\right)$ as a function of $\tk$ for $\mu=-2.0t$ and
$J=2.0t$ and three different approximant sizes $N_{a}$. The dashed
lines are a power-law fits with $\tkma=10^{-4}t$ with the corresponding
exponents $\alpha$ shown in the caption. Inset: Minimum value of
the Kondo temperature $\tkm$ a function of the inverse approximant
size. The dashed line is a power-law fit with $\tkm\propto N_{a}^{-0.69}$.}
\end{figure}

In this work, we consider different values of $N_{a}$ in order to
establish what happens for a true quasicrystal $\left(N_{a}\rightarrow\infty\right)$.
As we mention in the main text, for all approximants sizes $N_{a}$
we find a minimum Kondo temperature in the sample, $\tkm$. For the
smaller approximants, the six local environments of the octagonal
tiling (Fig. 1(b) of the main text) appear in a modest number of different
arrangements. In other words, their extended environment, including
next-nearest and further neighbors is limited. This leads not only
to an appreciable $\tkm$ but also to few distinct values of $\tk$.
As we increase the approximant size, these local environments appear
in further unique configurations leading to more and more values of
$\tk$ in the sample. Therefore, we expect the statistics of $\tk$
to improve with $N_{a}$, which can be clearly seen in Fig. \ref{fig:PTK_versusN},
where, for instance, the peak around $\tkt\sim10^{-3}t$ (which hardly
varies with $N_{a}$) becomes ever more well defined as the system
size increases. The most important, however, is the ubiquitous presence
of the power-law tail at low-$\tk$ in Fig. \ref{fig:PTK_versusN}
for all three approximant sizes with the same exponent (within error
bars). Moreover, it is also clear from Fig. \ref{fig:PTK_versusN}
that $\tkm$ is suppressed with increasing $N_{a}$. Indeed, in the
inset of Fig. \ref{fig:PTK_versusN} we find a power-law dependence
of $\tkm$ on $N_{a}$: $\tkm\propto N_{a}^{-0.69}$. Such power-law
finite-size scaling (with a nontrivial power) is precisely what one
expects in a critical state (in a conventional metal, for instance,
one would expect power-law finite-size scaling---as it is gapless---but
with integer powers).

Within our model, FL behavior is restored below $\tkm$ as all local
moments are then screened. Since the power-law distribution of Kondo
temperature $P\left(\tk\right)\propto\tk^{\alpha-1}$ emerges for
$\tk<\tkt$, we could expect, in principle, the NFL range to be constrained
to the interval $\tkm<T<\tkt$. However, as Fig. \ref{fig:PTK_versusN}
shows, $\tkm$ vanishes as $N_{a}$ increases while $\tkt$ remains
finite. We thus conclude that the NFL range actually extends down
to $T=0$ in a real quasicrystal. Therefore, our results suggest that
it is not their local structure, but the lack of long-distance periodicity
which induces robust NFL behavior in quasicrystals. %%%%% Since we are chiefly interested in the leading low- asymptotics (the value of the power), we can safely assume that finite-size effects are negligible in our current discussion. %%%%%

%%%%%%%%%%%%%%%%%%%%%%%%%%%%%%%%%%%%%%%%%%%%%%%%%%%%%%%%%%%%%%%%%%%%%%%

\section{Numerical calculation of the power-law exponent $\alpha$}

\begin{figure}[t]
\begin{centering}
\includegraphics[scale=0.28]{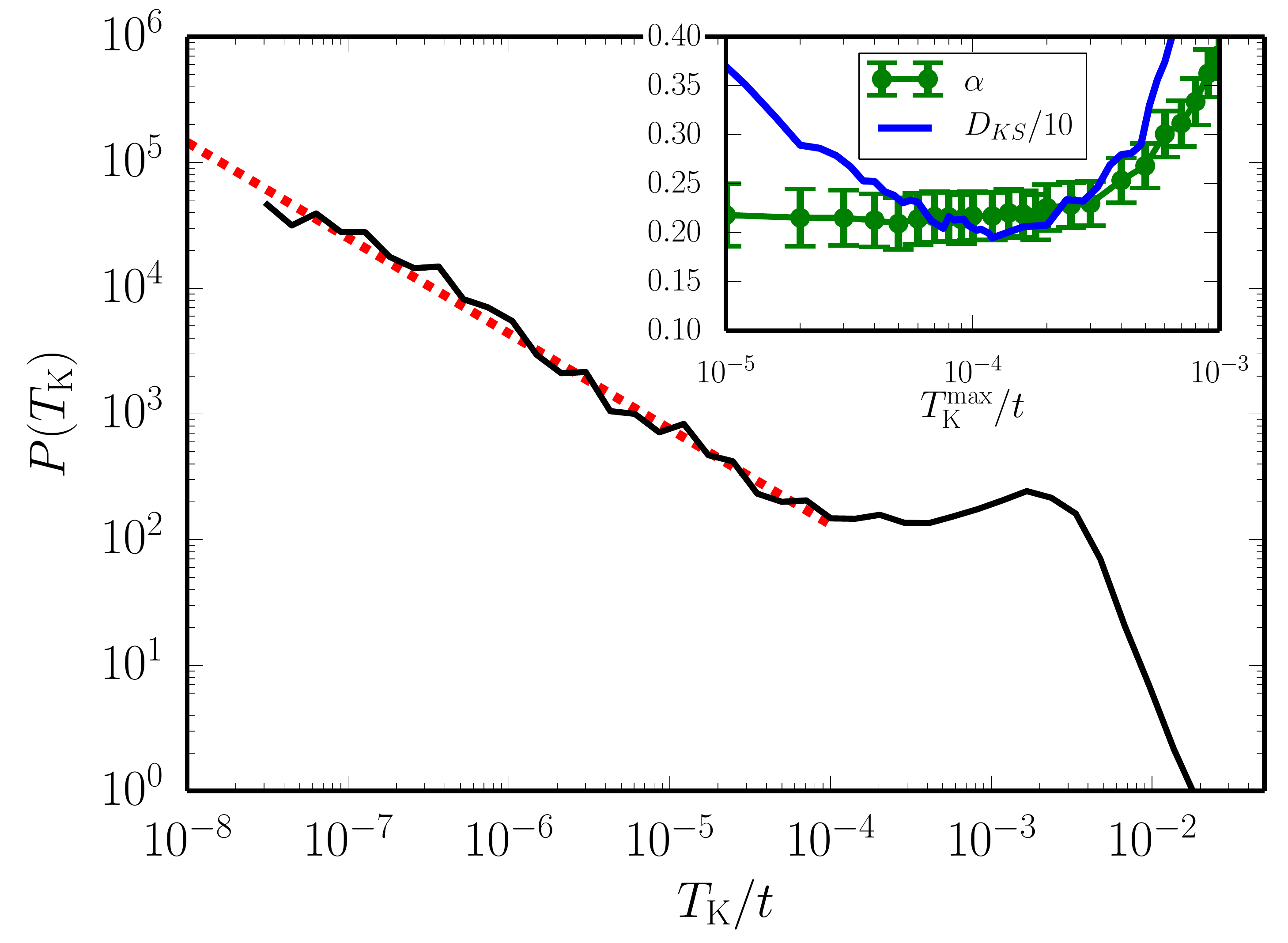}
\par\end{centering}

\caption{\label{fig:PTK_estimator} Octagonal tiling. Distribution of Kondo
temperatures $P\left(\tk\right)$ as a function of $\tk$ for $\mu=-2.0t$
and $J=2.0t$. The dashed line shows a power-law fit with $\alpha=-0.24$
and $\tkma=10^{-4}t$. Inset: The circles shows the value of $\alpha$,
as calculated from Eq. \eqref{eq:estimator}, whereas the full line
show the Kolmogorov-Smirnov (KS) test as a function of $\tkma$. Here
we considered the $N_{a}=1393$ approximant.}
\end{figure}

Here we address how we calculate the power-law exponent $\alpha$
governing the low-$\tk$ part of the distribution of Kondo temperatures.
The straightforward way is to plot $P\left(\tk\right)$ on a log-log
scale and then extract $(\alpha-1)$ as the slope of the resulting straight
line. While well defined, this procedure extracts $\alpha$ not from
the data itself, but from a given histogram. We complement the latter
procedure calculating $\alpha$ directly from the data as explained
in Ref.~\onlinecite{clauset09}.

Given a data set containing $n$ observations $\tk\le\tkma$, where
$\tkma$ is the largest value of the energy scale for which the power-law
distribution holds, the value of $\alpha$ that is most likely to
have generated our data is given by
\begin{eqnarray}
\alpha & = & \frac{n}{\sum_{i=1}^{n}\mbox{ln}\left[\tkma/\tk^{i}\right]},\label{eq:estimator}
\end{eqnarray}
with an error
\begin{eqnarray}
\sigma_{\alpha} & = & \alpha/\sqrt{n}.\label{eq:sigma_estimator}
\end{eqnarray}

In practice, however, the greatest source of error comes from not
choosing an optimal value for $\tkma$, which we dub $\tkmas$. We
then implement two procedures to estimate $\tkmas$.\cite{clauset09}
In the first one, we plot $\alpha\times\tkma$ and define $\tkmas$
as the point around which $\alpha$ is stable as we vary $\tkma$.
The second procedure follows the spirit of a chi-square test. The
idea is to investigate how well our data is fitted by a power-law
distribution. Since we are now dealing with distributions, we implement
the so-called Kolmogorov-Smirnov (KS) test.\cite{numerical_recipes}
The KS statistics $D_{KS}$ is defined as the maximum value of the
absolute difference between the two \emph{cumulative} distribution
functions. We then attempt to minimize $D_{KS}$ as a function of
$\tkma$.

In Fig. \ref{fig:PTK_estimator} we illustrate the discussion above.
In the main panel we show $P\left(\tk\right)$ on a log-log plot accompanied
by a power-law fit to its low-$\tk$ tail. In the fit displayed here,
we considered $\tkma=10^{-4}t$ and obtained $\alpha=0.24$. In the
inset we then show our two proposed tests to estimate $\tkmas$. We
see that the the $D_{KS}$ statistics has a minimum around $\tkma=10^{-4}t$
and that in this region $\alpha$ is essentially flat as a function
of $\tkma$, with a value of $\alpha=0.22\pm0.03$.

%%%%%%%%%%%%%%%%%%%%%%%%%%%%%%%%%%%%%%%%%%%%%%%%%%%%%%%%%%%%%%%%%%%%%%%

\section{Different tilings}

\begin{figure}[t]
\begin{centering}
\includegraphics[scale=0.27]{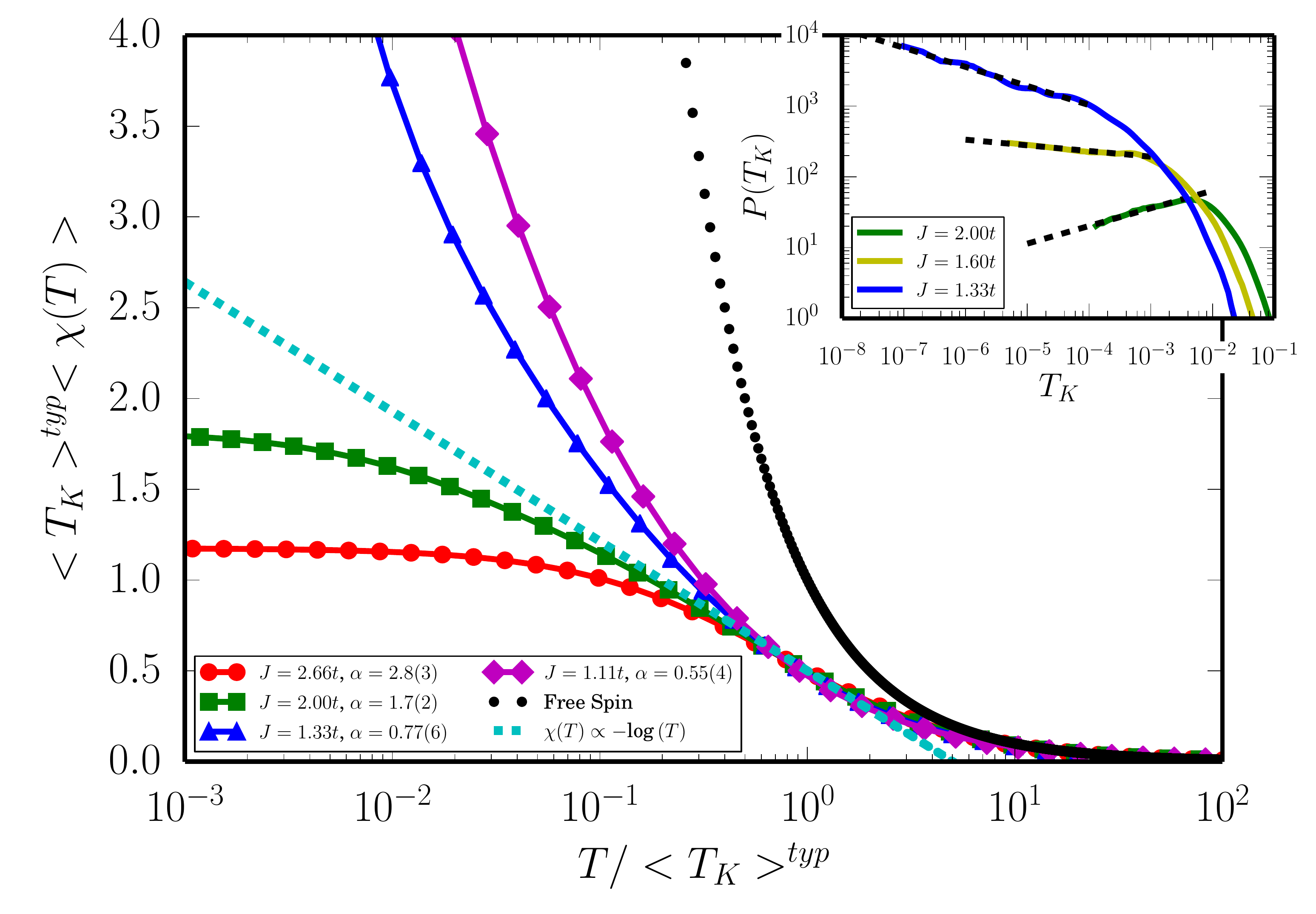}
\par\end{centering}

\caption{\label{fig:chiT_ico} Icosahedral tiling. Averaged value of the impurity
susceptibility $\left\langle \chi\left(T\right)\right\rangle $ times
the typical value of the Kondo temperature $\tkt$ as a function of
the temperature $T$ normalized by $\tkt$ for four values of the
Kondo coupling $J$ on a semi-log scale. For completeness, we show
both the free spin and the $\chi\propto-\mbox{log}\left(T\right)$
$\left(\alpha=0\right)$ curves. Inset: Distribution of the local
Kondo temperatures $P\left(\tk\right)$ as a function of $\tk$ on
a log-log scale for three values of the Kondo coupling $J$. $J$
increases from the top to the bottom curve. At low $\tk$ this distribution
acquires a power-law form $P\left(\tk\right)\sim\tk^{\alpha-1}$.
The power-law exponent $\alpha$ continuously varies with the coupling
$J$ and for $\alpha<1$ we have a singular distribution. Here we
considered $\mu=-3.0t$, $N_{a}=576$, and $N_{\phi}=512$.}
\end{figure}

So far we have investigated the single-impurity Kondo effect in the octagonal tiling, a $2D$
quasicrystal, while the original motivation came from experiments on a $3D$ heavy
fermion quasicrystal $\aualyb$,\cite{deguchi12,watanuki12}.
In the following we show that this difference in spatial dimensionality does not change
the qualitative behavior of Kondo impurities.
First we note that the importance of spatial dimensionality in determining
the statistics of wavefunction amplitudes (e.g. the local density
of states statistics) is well known in disordered systems.\cite{anderson58}
For this problem, $2D$ and $3D$ are significantly different because
$2D$ is the lower critical dimension for Anderson localization.\cite{gang4}
Second, however, it is known that some models for low-dimensional
(even $1D$) quasicrystals can support extended or pseudo-extended
electronic states, and even a sharp Anderson-like transition and a
mobility edge.\cite{luck85} In this sense, $2D$ is most likely {\em not}
the lower critical dimension for wavefunction localization in quasicrystals.
Hence, there should not be a significant qualitative difference between
electronic quasicrystalline states in $2D$ and $3D$.\cite{grimm03}
Therefore we expect that our $2D$ results capture the key
effects of the quasicrystalline wavefunctions on the Kondo effect in general,
and that our conclusions should remain valid in $3D$ thus providing
a robust and general scenario for the emergence of NFL behavior in
quasicrystals.

To support the claim that our results are general and applicable to different tilings (even to $3D$ quasicrystals), we have studied the Kondo problem in
the $3D$ icosahedral tiling.\cite{zijlstra00} This tiling possesses
$7$ distinct local environments with coordination number $z=4,\cdots,9,\mbox{ and },12$.
The average coordination number is $6$ and the bandwidth is comparable
to that of the simple cubic lattice. Sample results are presented
in Fig. \ref{fig:chiT_ico}. As in the octagonal tiling, we obtain a
power-law distribution of Kondo temperatures, $P\left(\tk\right)\sim\tk^{\alpha-1}$,
and the corresponding NFL behavior for $\alpha<1$. As discussed in
the main text, the essential ingredient for this behavior is the unanticipated
Gaussian form of the local $c$-electron cavity function fluctuations
at the Fermi level, $\delta\Delta_{c}^{\prime}$, and not any special
lattice symmetry or dimension.

In conclusion, the striking similarity of our results for the octagonal and icosahedral
tilings shows that NFL behavior from dilute Kondo impurities in quasicrystals is robust
and serves as a starting point to understand quasicrystalline Kondo lattices, 
in order to connect to the recently observed NFL behavior in the $3D$ heavy
fermion quasicrystal $\aualyb$.\cite{deguchi12,watanuki12}

%%%%%%%%%%%%%%%%%%%%%%%%%%%%%%%%%%%%%%%%%%%%%%%%%%%%%%%%%%%%%%%%%%%%%%%